\documentclass[lettersize,journal]{IEEEtran}
\ifCLASSOPTIONcompsoc
  \usepackage[nocompress]{cite}
\else
  \usepackage{cite}
\fi

\usepackage{cite}
\usepackage{amsmath,amssymb,amsfonts}
\usepackage{algorithmic}
\usepackage{graphicx}
\usepackage{textcomp}
\usepackage[hyphens]{url}
\usepackage{listings}
\usepackage{float}
\usepackage{pifont}
\usepackage{array}
\usepackage{stfloats}
\usepackage{verbatim}

\usepackage[caption=false,font=normalsize,labelfont=sf,textfont=sf]{subfig}

\usepackage[breaklinks,colorlinks]{hyperref}
\usepackage[usenames,dvipsnames]{xcolor}
\hypersetup{citecolor=blue,linkcolor=blue}
\usepackage{amsmath,amsopn,amssymb}
\usepackage{endnotes,microtype,xspace,graphicx,fancyvrb,multirow,diagbox}
\usepackage{booktabs}
\usepackage[T1]{fontenc}
\usepackage{fancyhdr,lastpage}
\usepackage{enumitem}
\usepackage[labelfont=bf,font=small,skip=5pt]{caption}
\usepackage{pifont}

\usepackage{fp}
\usepackage{siunitx}

\usepackage{balance}

\usepackage{soul}

\def\BibTeX{{\rm B\kern-.05em{\sc i\kern-.025em b}\kern-.08em T\kern-.1667em\lower.7ex\hbox{E}\kern-.125emX}}

\usepackage{authblk}

\usepackage{CJKutf8}

\usepackage{centernot}
\usepackage{paracol}
\usepackage{multicol}
\usepackage{adjustbox}

\usepackage[linewidth=1pt]{mdframed}
\usepackage{bbding}
\usepackage{orcidlink}
\usepackage{subfloat}

\definecolor{mygreen}{rgb}{0,0.6,0}
\definecolor{mygray}{rgb}{0.5,0.5,0.5}
\definecolor{mymauve}{rgb}{0.58,0,0.82}
\lstdefinelanguage{myC}{
	backgroundcolor=\color{white},      
	basicstyle=\footnotesize\ttfamily,
	columns=fullflexible,
	tabsize=4,
	breaklines=true,               
	captionpos=b,                  
	commentstyle=\color{mygreen},  
	escapeinside={(*}{*)},        
	keywordstyle=\color{blue},     
	morekeywords={foreach, int, char, bool, struct, ulong, return, asm, if, else, in, delete, void, update, with, then, unsigned, long, define, uint8},
	stringstyle=\color{mymauve}\ttfamily,  
	frame=single,
	rulesepcolor=\color{red!20!green!20!blue!20},
	numbersep=5pt, 
	numberstyle=\tiny,
	numbers=left, 
	morecomment=[l]//,%
	morecomment=[s]{/*}{*/},%
	morestring=[b]",%
	morestring=[b]',%
	xleftmargin = 1em,
}

\newenvironment{packeditemize}{
\begin{list}{$\bullet$}{
\setlength{\labelwidth}{8pt}
\setlength{\itemsep}{0pt}
\setlength{\leftmargin}{\labelwidth}
\addtolength{\leftmargin}{\labelsep}
\setlength{\parindent}{0pt}
\setlength{\listparindent}{\parindent}
\setlength{\parsep}{2pt}
\setlength{\topsep}{3pt}}}{\end{list}}

\newcommand{\bheading}[1]{{\vspace{4pt}\noindent{\textbf{#1}}}}
\newcommand{\tabincell}[2]{\begin{tabular}{@{}#1@{}}#2\end{tabular}}

\newcommand{\sys}{\mbox{\textsc{SpecWands}}\xspace}

\newcommand{\indep}{\perp \!\!\! \perp}
\newcommand{\notiff}{%
  \mathrel{{\ooalign{\hidewidth$\not\phantom{"}$\hidewidth\cr$\iff$}}}}

\usepackage{tikz}

\begin{document}

\title{\sys: An Efficient Priority-based Scheduler Against Speculation Contention Attacks}


\author{
    Bowen Tang$^{\orcidlink{0000-0002-5845-4684}}$,
    Chenggang Wu$^{\orcidlink{0000-0003-1777-8110}}$, 
    Pen-Chung Yew$^{\orcidlink{0000-0001-9653-8777}}$,~\IEEEmembership{Fellow, IEEE},
    Yinqian Zhang$^{\orcidlink{0000-0002-7585-1075}}$, 
    Mengyao Xie$^{\orcidlink{0000-0002-8511-1118}}$,\\
    \vspace{-1.2em}
    Yuanming Lai$^{\orcidlink{0000-0001-5885-0858}}$,
    Yan Kang$^{\orcidlink{0000-0002-3439-551X}}$,
    Wei Wang$^{\orcidlink{0000-0003-1585-8731}}$,
    Qiang Wei$^{\orcidlink{0000-0002-7207-6691}}$,
    Zhe Wang$^{\orcidlink{0000-0003-4719-1804}}$
    
    \IEEEcompsocitemizethanks{
        This research was supported by the National Natural Science Foundation of China (NSFC) under Grant 61902374 and U1736208, and Innovation Funding of ICT, CAS under Grant No.E161040. \emph{(Corresponding Author: Zhe Wang.)}

        Bowen Tang, Chenggang Wu, Mengyao Xie, Yuanming Lai, Yan Kang, Wei Wang, Zhe Wang are with SKLP, Institute of Computing Technology, CAS, Beijing 100190, China, and also with University of Chinese Academy of Sciences, Beijing 100049, China. Chenggang Wu and Zhe Wang are also with Zhongguancun Laboratory, Beijing, China. Email: \{tangbowen, wucg, xiemengyao, laiyuanming, kangyan, wangwei2021, wangzhe12\}@ict.ac.cn
        
        Pen-Chuang Yew is with Computer Science \& Engineering Department of the University of Minnesota-Twin Cities, MN 55455, USA. Email: yew@umn.edu
        
        Yinqian Zhang is with Department of Computer Science and Engineering of the Southern University of Science and Technology, Shenzhen 518055, China. Email: yinqianz@acm.org
        
        Qiang Wei is with State Key Laboratory of Mathematical Engineering and Advanced Computing, ZhengZhou 450000, China. Email: prof\_weiqiang@163.com\\
    }

}


\maketitle
\begin{abstract}
Transient Execution Attacks (TEAs) have gradually become a major security threat to modern high-performance processors. They exploit the vulnerability of speculative execution to illegally access private data, and transmit them through timing-based covert channels. While new vulnerabilities are discovered continuously, the covert channels can be categorised to two types:
1) \emph{Persistent Type}, in which covert channels are based on the layout changes of buffering, e.g. through caches or TLBs; 2) \emph{Volatile Type}, in which covert channels are based on the contention of sharing resources, e.g. through execution units or issuing ports. The defenses against the  persistent-type covert channels have been well addressed, while those for the volatile-type are still rather inadequate. 
Existing mitigation schemes for the volatile type such as \emph{Speculative Compression} and \emph{Time-Division-Multiplexing} will introduce significant overhead due to the need to stall the pipeline or to disallow resource sharing. In this paper, we look into such attacks and defenses with a new perspective,
and propose a scheduling-based mitigation scheme, called \sys. It consists of three priority-based scheduling policies to prevent an attacker from transmitting the secret in different contention situations. \sys not only can defend against both inter-thread and intra-thread based attacks, but also can keep most of the performance benefit from speculative execution and resource-sharing. 
We evaluate its runtime overhead on SPEC 2017 benchmarks and realistic programs. The experimental results show that \sys has a significant performance advantage over the other two representative schemes.
\end{abstract}


\section{Introduction}\label{sec:intro}
\IEEEPARstart{S}{imultaneous} multi-threading (SMT), also known as hyper-threading, has become an important feature on modern high-performance processors. It allows multiple threads to run simultaneously on a physical core and share the resources in the instruction pipeline to cover the slack caused by the stalled threads, thereby improving the efficiency and throughput of the pipeline.
However, due to the resource sharing on the pipeline, SMT can introduce new security vulnerabilities. Multiple threads may compete for shared resources and interfere with each other's execution under the common \emph{first-come-first-served} (FCFS) policy. 
If an attacker can control one thread, he/she can figure out the execution state of other threads according to the difference in its execution time and then infer some private information, which is the so-called SMT contention-based side channel \cite{fpTiming, SIMDTiming, portTiming}.

Such attacks can be mitigated by scheduling mutually distrusting threads on different physical cores. However, the recent \emph{transient execution attacks} (TEAs) \cite{TEAEvaluation, foreshadow, mds1, LVI, spectreSTL, swapgs, netspectre, spectreInterference, meltdown0, spectre0}, such as the well-known Meltdown and Spectre, cannot be defended using such an approach. It is because an attacker can spawn and control multiple threads in some attack scenarios. The system cannot distinguish which ones are malicious when resource contention occurs.

Take an example from the work in ~\cite{spectreSmother}.
It installs a malicious plug-in running on a sandboxed browser to launch such an attack.
Assume two threads are both created by the plug-in and run on a physical core with SMT.
One of them, called \emph{Trojan}, leverages speculative execution to access a secret outside the sandbox. 
Trojan thread then issues a burst of requests to keep the resource busy if the bit to be transmitted is "1", and leaves the resource idle if it is "0" instead. The other thread, called \emph{Spy}, then attempts to acquire the same resource and infer the secret value based on whether the resource is busy or not by measuring its acquisition time. In such an attack scenario, the resource contention is used as a covert channel to transmit the illegally obtained data from speculative accesses.

To block such covert channels, researchers have proposed a variety of defenses. One well-known approach is \emph{Speculation Compression}~\cite{STT, NDA, SpecShield},
which delays the speculative data to be propagated in the pipeline.
That is, if an operand of an instruction comes from a speculative instruction, it will not be issued until the speculative instruction it depends on has become non-speculative. 
This approach prevents the data of a speculative instruction from being transmitted via the covert channel. However, it will obstruct the speculative execution (e.g. from branch prediction) that has been the cornerstone of a modern CPU to improve its performance.
Based on our own simulation results, its performance hit can be as high as 15\% on a typical CPU with SMT.
The other approach is to partition the resources in the spacial or temporal dimensions, such as \emph{Time-Division-Multiplexing} (TDM), to avert inter-thread interference~\cite{SMT-COP, SMT-COPImproved, secSMT}. 
Although this approach doesn't obstruct the speculative execution, it contradicts the original intention of resource sharing using SMT, and will also incur non-negligible performance overhead (e.g. more than 12\% overhead according to our evaluation).

Besides the well-known resource contention in SMT, researchers have found the resource contention also exists in the single thread scenario, and it can be abused as a new convert channel~\cite{SpectreRewind, woSMT}.
Such attacks exploit the contention caused by multiple instructions scheduled in the same scheduling window within a thread. These instructions have no dependence among them, and can be issued simultaneously in the pipeline that supports \emph{multi-issuing} and \emph{out-of-order} execution. 
The attacker can use the contention at the issuing port as a covert channel.
Moreover, he/she only needs to launch one single thread without the need for inter-thread synchronization.
This kind of attacks, which leverage intra-thread covert channels, thus have a higher probability of success. 
TDM-based defense approaches are basically ineffective and are difficult to  harden for this scenario.

In order to defend against the TEAs that exploit resource contention as a covert channel, which we call \emph{speculative contention attacks} (SCAs), while minimizing the performance overhead, we proposed a secure scheduling scheme for shared resources, called \sys. 
The main idea behind \sys is to batching multiple operations into groups, and ensure that each group can share a resource without any interference from other groups in the same or different threads.
To apply such an idea to different scenarios, we use the following three priority-based scheduling policies.



\begin{packeditemize}
\item \bheading{NOP: Non-speculative Operations have a higher Priority.} 
\sys assigns a higher priority to non-speculative operations. They not only can be scheduled ahead of all other speculative operations, but also allowed to preempt the speculative operations already occupying the resource. 
This policy can be used as a guideline for delimit the boundaries of each group, i.e. each non-speculative operation is the header of each group.


\item \bheading{LOP: Last-owner-thread's Operations have a higher Priority.} 
When there are multiple speculative operations from different threads, \sys assigns a higher priority to the speculative operations whose owner thread used the resource most recently. This policy further clusters the speculative operations into groups based on their owner threads, i.e. their group header non-speculative operations.


\item \bheading{EOP: Earlier Operations have a higher Priority.} 
Within the same SMT thread, when the operations that request a resource are all speculative, \sys assigns a higher and preemptive priority (similar to the NOP policy) to the operations that are issued earlier in program order. Different from NOP and LOP policies, its purpose is to prevent the backward contention within a group, i.e., to prevent a later operation from blocking an earlier operation in the group  due to multi-issuing and out-of-order execution. 

\end{packeditemize}


The above policies not only maintain the security principle of \emph{Speculative Non-Interference} (SNI)\cite{SNI}, but also facilitate the temporal locality of the shared resource usage.
From our simulation results, \sys only introduces 1\% and 5\% performance overhead on realistic programs and SPEC 2017 benchmarks, respectively, which are much lower than those in STT \cite{STT} and SMT-COP \cite{SMT-COP}, the representative work of the other two defenses.

To summarize, this paper makes the following contributions.
\begin{enumerate}
    \item We examine the deficiencies in existing defense approaches against SCAs, such as speculation compression and TDM, and propose a novel mitigation scheme to reduce most delay between operations within the group, while blocking potential cover channels created in both inter-thread and intra-thread modes.
    
    
    \item We present a hardened instruction scheduler practicing above scheme, called \sys. It consists of three priority-based scheduling policies for different contention situations to divide the instructions into groups and ensure their security. Moreover, we analyze the performance impact of each policy based on the distribution of different contention situations.
    
    \item We evaluate the performance overhead and power consumption of \sys via detailed simulations. Compared with the other two representative defenses\cite{STT,SMT-COP}, \sys has lower overheads and power consumption for realistic programs and SPEC 2017 benchmarks.
\end{enumerate}

\begin{figure}[!t]
	\centering
	\includegraphics[width=\columnwidth]{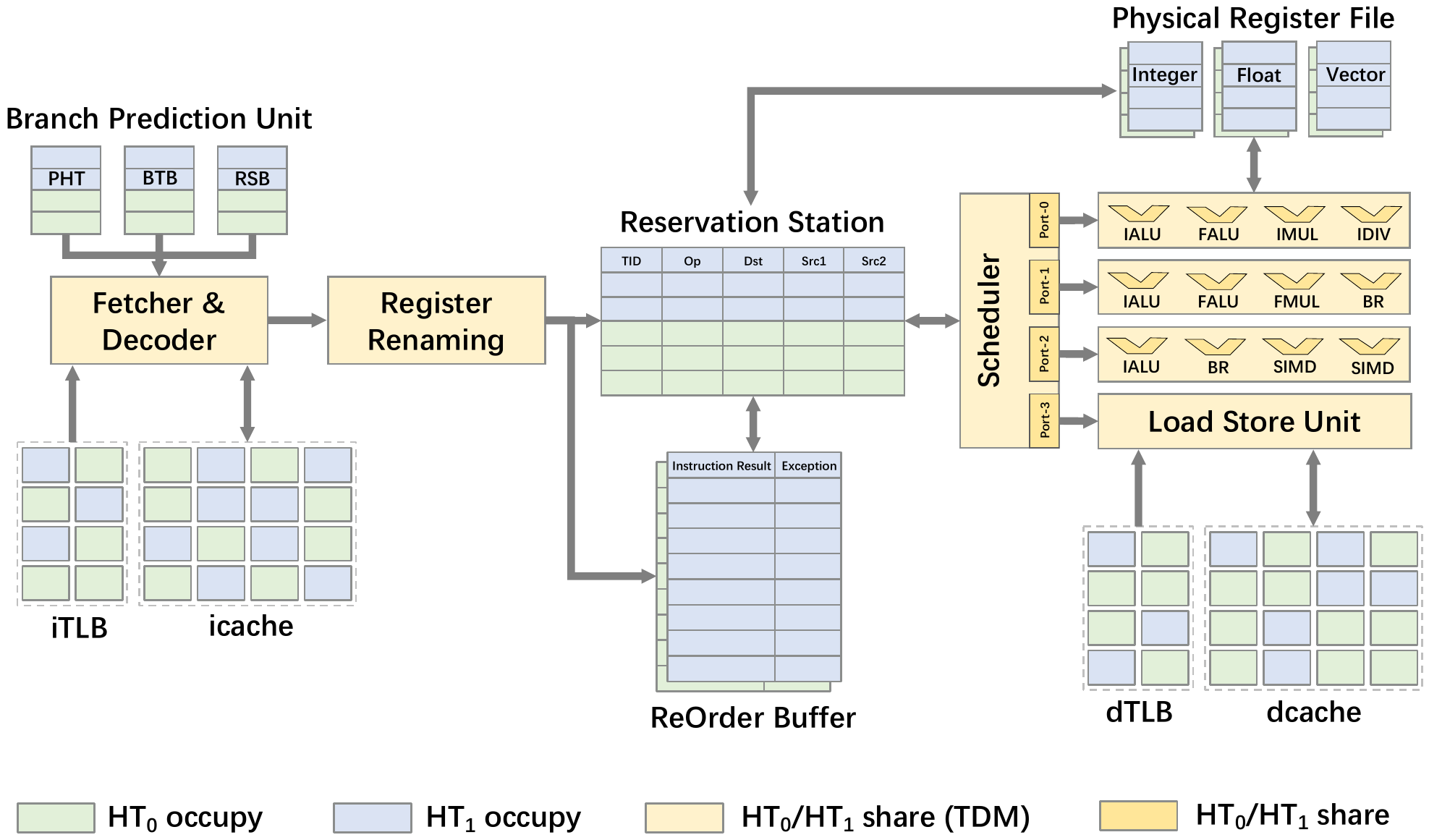}
	\caption{The microarchitecture of a typical high-performance CPU that supports \emph{speculative}, \emph{out-of-order} execution with a 2-context SMT.}
	\label{fig:SMT}
	\vspace{-1.5em}
\end{figure}

\vspace{-0.5em}
\section{Background}\label{sec:bkg}
\subsection{Modern CPU Pipeline and SMT}
\autoref{fig:SMT} shows a typical microarchitecture of a modern high performance CPU. The branch prediction unit, which consists of PHT/BTB/RSB, facilitates the control-flow speculation, while renaming-based instruction scheduling and a ROB with a squashing mechanism support the out-of-order program execution. These techniques optimize the use of instruction level parallelism (ILP) and can improve the processor performance quite substantially. To further improve the throughput of an instruction pipeline, switching from a thread that encounters a long-delay cache miss to another thread ready for execution can hide the memory latency and improve resource utilization from another view. This approach eventually led to the design of \emph{simultaneous multi-threading} (SMT)\cite{smt0}. 

In an SMT processor, multiple \emph{hardware-supported threads} (denoted as HTs in the rest of the paper) are running on the pipeline concurrently and share most of its critical resources. The storage resources, such as physical register file (PRF), ROB and PHT/BTB/RSB are shared among HTs by physical partitioning them or using tag-based partitioning schemes. It allows the states of multiple HTs to be maintained simultaneously without using context switching. Some computational components, such as the instruction fetcher, decoder and register-renaming unit, are shared among HTs using a time-division-multiplexing (TDM) scheme. For L1-cache/TLBs, execution units and issuing ports, they are fully shared by all HTs to maximize their utilization. Earlier studies have shown that, compared with a single-threaded processor, the performance improvement brought on by a 2-context SMT processor can be up to 30\% with negligible hardware cost~\cite{smt0}. Hence, SMT has been widely used in PCs and servers.


\vspace{-1em}
\subsection{Side Channel Attacks on SMT Processors}
While resources sharing can bring performance benefits, it also exposes the SMT processors to potential side channel attacks. For the shared storage components, such as L1-caches/TLBs, the interference among sharing HTs can cause a subsequent access to be a hit or a miss. An attacker can exploit this side effect by measuring the difference in its access time, and infer some private information such as the encryption key or keyboard strokes through the access traces\cite{channelSurvey}. Because the placement and the measurement of the data layout in the side channel can be done asynchronously, such channels are called \emph{persistent} channels. For shared computational components, such as issuing ports\cite{portTiming} and execution units\cite{fpTiming, SIMDTiming}, the contention among threads can prolong the execution time of each HT, which can also be perceived by the attacker to reason about the private information related to the execution flow. Such side channels are referred to as \emph{volatile} channels.

\vspace{-1em}
\subsection{Transient Execution Attacks and Covert Channels}\label{subsec:covert}
\begin{figure}[!t]
	\centering
	\includegraphics[width=\columnwidth]{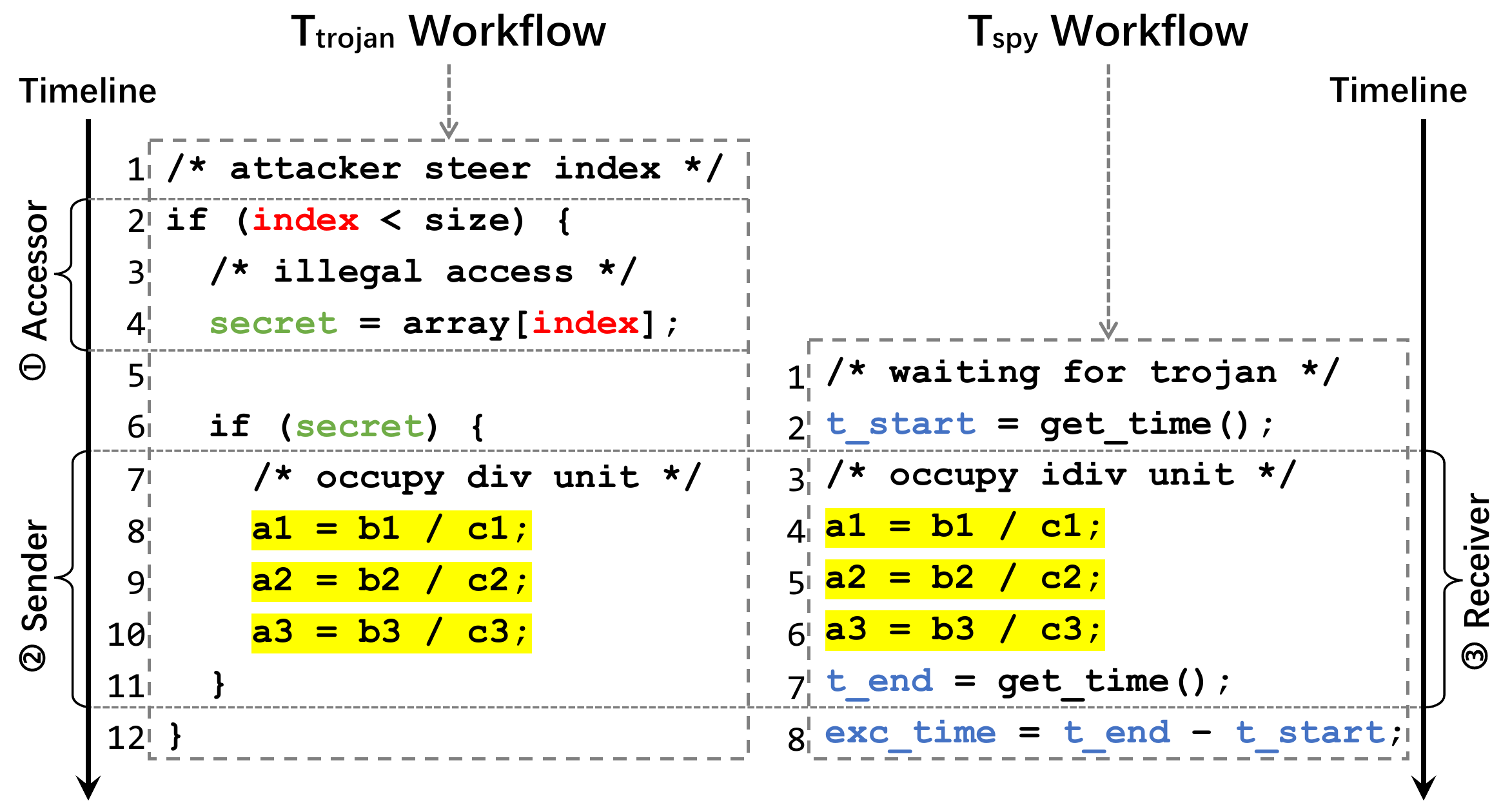}
	\caption{The PoC of an inter-thread SCA that exploits Spectre-V1 vulnerability. The "index" can be steered by an attacker, and the "secret" is the target to be leaked. The highlighted code can occupy an integer division unit and its corresponding issuing port.}
	\label{fig:poc_inter}
	\vspace{-1.5em}
\end{figure}

Side channel attacks have not attracted attention of processor designers until the emergence of \emph{transient execution attacks} (TEAs). The core logic of a TEA has three main steps: \emph{Accessor, Sender} and \emph{Receiver}\cite{TEAEvaluation, TEASurvey}. In the accessor step, the attacker illegally accesses the secret data through a staged speculative execution. 
There are two ways the attacker can set up the speculative attacks.
In \emph{Spectre-type} attacks, the attacker exploits hardware branch predictors on control flow transfer or memory disambiguation to bypass intended software defense codes, such as bound checking\cite{spectre0}, data cleaning\cite{spectreSTL} or stack pivoting\cite{swapgs}.
While in \emph{Meltdown-type} attacks, the attacker exploits the bugs of access permission protection to break hardware isolation between domains, such as User\_Space and Kernel\_Space\cite{}, GuestVM and Hypervisor, or SGX\_Enclave and Untrusted\_Software\_Stack \cite{meltdown0, foreshadow}.

The attacker can then use the \emph{sender} step and the \emph{receiver} step to transmit the illegally accessed data through a covert channel. 
A covert channel works very similarly to that of a side channel. The only difference is that the sender and the receiver in a covert channel are both manipulated by the attacker.
While in a side channel, the victim process is the sender and the receiver are controlled by the attacker.
In this paper, we use the term \emph{Speculative Contention Attacks} (SCAs) to describe the TEAs that are based on resource contention, i.e. the attacks that use \emph{volatile covert channels}.
Furthermore, we classify the existing SCAs into two categories.

\bheading{Inter-thread SCAs.} \autoref{fig:poc_inter} shows a PoC code of the Spectre-V1 attack that exploits the resource contention on the issuing port and the execution unit on an SMT processor. In this attack, $HT_{Trojan}$ and $HT_{Spy}$ are both controlled by the attacker, with $HT_{Trojan}$ acting as both the accessor and the sender, and $HT_{Spy}$ acting as the receiver. Firstly, $HT_{Trojan}$ bypasses the bound check by manipulating the branch predictor at line 2, and reads the secret by out-of-bounds access at line 4 during the speculative execution. Then, $HT_{Trojan}$ controls the execution of the subsequent division instructions according to the value of secret. 
Assume the secret has only one bit.
If it is 1, the division will be executed. Meanwhile, $HT_{Spy}$ will also perform division operations and measuring the time. If the time is shorter, it can infer that no contention has occurred and the secret is 0; Otherwise, the secret is 1.

\begin{figure}[!t]
	\centering
	\includegraphics[width=0.95\columnwidth]{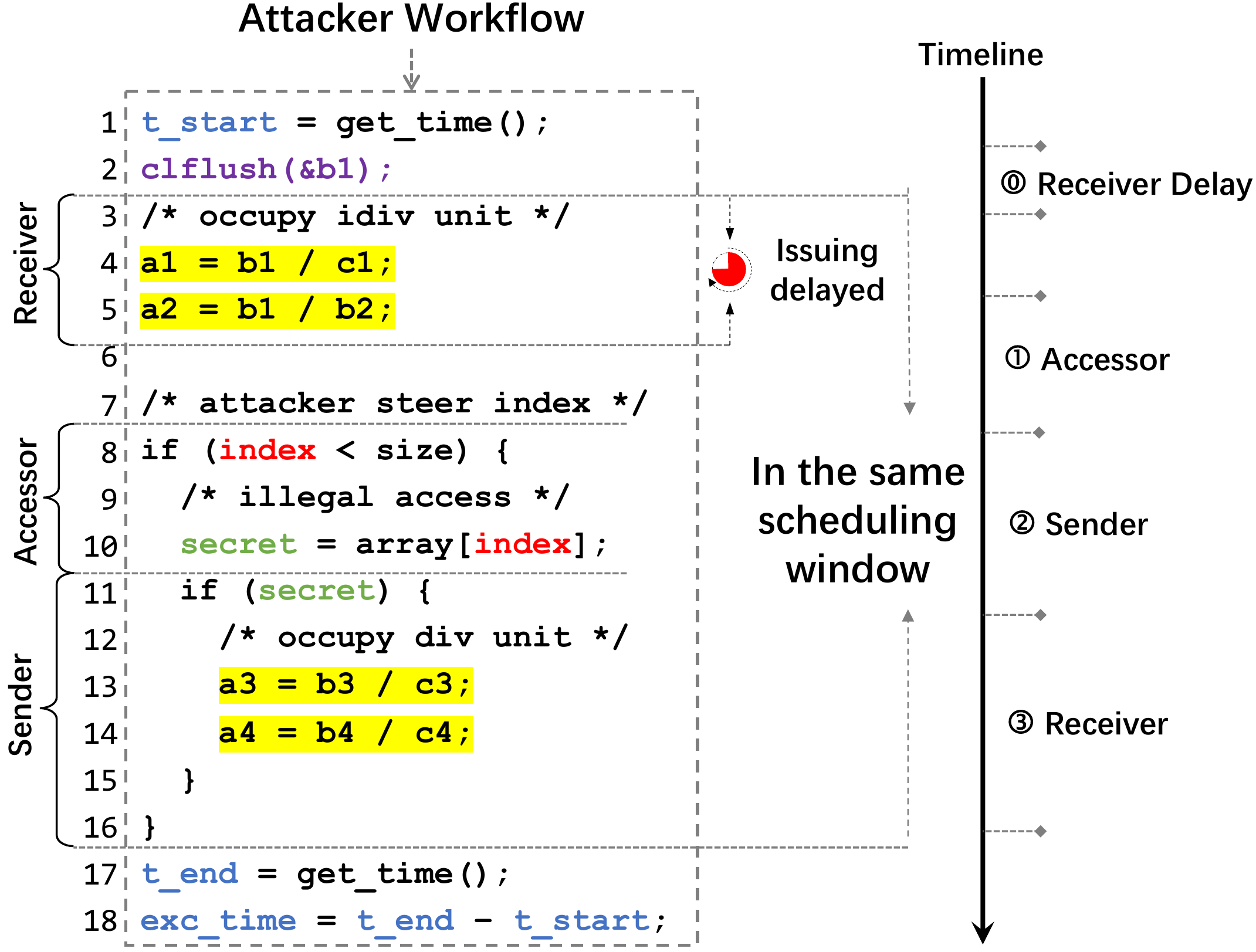}
	\caption{The PoC of an intra-thread SCA, also called \emph{SpecRewind} Attack, which also exploits Spectre-V1 vulnerability and leverages the intra-thread contention as covert channel.}
	\label{fig:poc_intra}
	\vspace{-1.5em}
\end{figure}

\bheading{Intra-thread SCAs.} 
 The above \emph{inter-thread} SCAs can be extended and carried out in the same HT, i.e. an \emph{intra-thread} SCA. Some researchers have shown that, on a superscalar processor, multiple independent instructions in the same scheduling window can compete for the resources and thus can be used to form a covert channel\cite{SpectreRewind, woSMT}. As \autoref{fig:poc_intra} shows, the attacker combines the three steps in the same HT, and delays the receiver step (via line 2) to make it executed simultaneously with the sender step. This attack is also known as \emph{SpecRewind} Attack\cite{SpectreRewind}.

\vspace{-0.5em}
\subsection{Existing Mitigations}\label{subsec:mitigations}
In most cases, mitigations for the attacks via persistent channels have been well established. They include domain partitioning\cite{KPart, DAWG}, index/replacement randomization\cite{RPcache, scatterCache}, and footprint-based detection \cite{RTDetect, MLDetect}. In particular, for TEAs through persistent covert channels, schemes that use \emph{Invisible Speculation}\cite{InvisiSpec, CleanupSpec, Muontrap, SpecBox} extends the squashing mechanism in the pipeline to cleanup and rollback the side effects in the cache memory for the mis-predicted speculative execution with modest performance and hardware overhead. 

However, these methods cannot be generalized and applied to volatile channels. 
Currently, the only secure solution is to disable SMT for security-sensitive HTs\cite{DDM}, or scheduling mutually untrusted threads on different physical cores\cite{partialSMT, contentionAware}. However, this approach only works against traditional side-channel attacks when the protected target, such as the thread executing a cryptographic computation or code inside a secure enclave, can be identified by the programmer before running. 
But for SCAs, any thread with vulnerable speculative code can be exploited by an attacker. It's hard to distinguish an $HT_{Trojan}$ from an $HT_{Spy}$ as they are often from the same user group and supposed to be trustworthy.

Another solution is to use \emph{Timing-Division-Multiplexing} (TDM) scheme on the shared resources\cite{SMT-COP}. It allocates time slices for each HT to avoid resource contention.
However, this approach violates the original intent of SMT to share unused resources when available. It can incur significant performance overhead due to its rigidity in resource sharing (more than 12\% in our evaluation). 
Although, some approaches try to adaptively allocate time slices to improve resource utilization \cite{SMT-COPImproved}, the adaptive measure can inevitably be used to become another potential covert channel.
Other schemes \cite{secSMT, underDome} adopt asymmetric allocation strategies for threads with different trust levels to ensure that highly trusted threads can obtain more time slices. 
However, the trust level needs to be specified by the programmer through some annotations.
So, it comes back to the earlier question of "which threads can we trust?".

Another type of defenses against SCAs is \emph{Speculation Compression}\cite{STT, NDA, SpecShield}. Instead of blocking covert channels, 
it disallows data or other potential microarchitectural side effects from propagating to the downstream instructions by stalling/blocking some dependent instructions during the speculative execution until the execution has reached some safe points. The advantage of such schemes is that they can block both persistent and volatile covert channels of TEAs comprehensively.
But, due to the use of stalling and blocking of the dependent instructions, they often incur a significant performance overhead (up to 15\% according to our evaluation).
\begin{table}[t]
  \centering
  \caption{The vulnerability and access capability of existing SCAs. \textbf{AC}: whether it can access the data cross hardware protection domains.}
  \resizebox{\columnwidth}{!}{
    \begin{tabular}{|l|l|l|l|}
    \hline
    \hline
        \textbf{Attack} & \textbf{Vunerability} & \textbf{AC} \\
    \hline
    \hline
        Spectre-PHT/BTB/RSB\cite{spectre0} & Control Flow Prediction & \XSolidBrush \\
    \hline
        Speculative Store Bypass\cite{spectreSTL} & Memory Disambiguiation & \XSolidBrush \\
    \hline
        Speculative Load Disorder\cite{spectreLD} & Memory Order Speculation & \XSolidBrush \\
    \hline
        SWAPGS\cite{swapgs} & Out-of-Order Execution & \Checkmark \\
    \hline
        Rogue System Register Read\cite{RSRR} & Buggy \#NM Exception Handler & \Checkmark \\
    \hline
        Meltdown\cite{meltdown0,TEAEvaluation} & Buggy \#GP Exception Handler & \Checkmark \\
    \hline
        L1TF (Foreshadow)\cite{foreshadow, TEAEvaluation} & Buggy Terminate Fault Handler & \Checkmark \\
    \hline
        MDS\cite{mds1}, LVI\cite{LVI} & Buggy Assist/Abort Data Forward & \Checkmark \\
    \hline
    \hline
    \end{tabular}
  }
  \label{tab:capability}
  \vspace{-1.5em}
\end{table}%

\vspace{-0.5em}
\section{Threat Model}\label{sec:threat}
\autoref{tab:capability} lists the vulnerabilities exploited by existing SCAs, as well as the capabilities for their illegal accesses. We assume a powerful attacker can launch any SCA listed in \autoref{tab:capability} within or across domains. For example, the attacker can inject an attack code through malicious Javascript scripts, or malicious browser plug-ins. The code can exploit Spectre-PHT vulnerability\cite{spectre0} to bypass the bound check of browser's sandbox and access some private keys and cookies.
Or, the attacker can exploit Meltdown vulnerability\cite{meltdown0} by crossing the hardware domain of the kernel and accessing some critical data structures. 
The attacker can then transmit the stolen secrets through a covert channel based on resource contention as shown in \autoref{fig:poc_inter} and \autoref{fig:poc_intra}. Furthermore, the attacker can launch a malicious virtual machine (VM) in the cloud and exploit Foreshadow vulnerability\cite{foreshadow} to access the data in other victim VMs residing on the same physical core. The malicious VM can also transfer the stolen data through the fabricated contention covert channel.

    
    

\bheading{Out-of-Scope.} We do not consider TEAs through persistent covert channels, e.g. cache or TLB. As explained in section~\ref{subsec:mitigations}, existing defenses are effective and efficient to cover these attacks, and they are orthogonal to our scheme. We also exclude non-transient side channel attacks because of their limited threat. But in \autoref{subsec:mitigations}, we still introduce some effective defenses against those attacks.


\section{Design of \sys}\label{sec:design}
\subsection{Overview}
\begin{figure}[!t]
	\centering
	\includegraphics[width=\columnwidth]{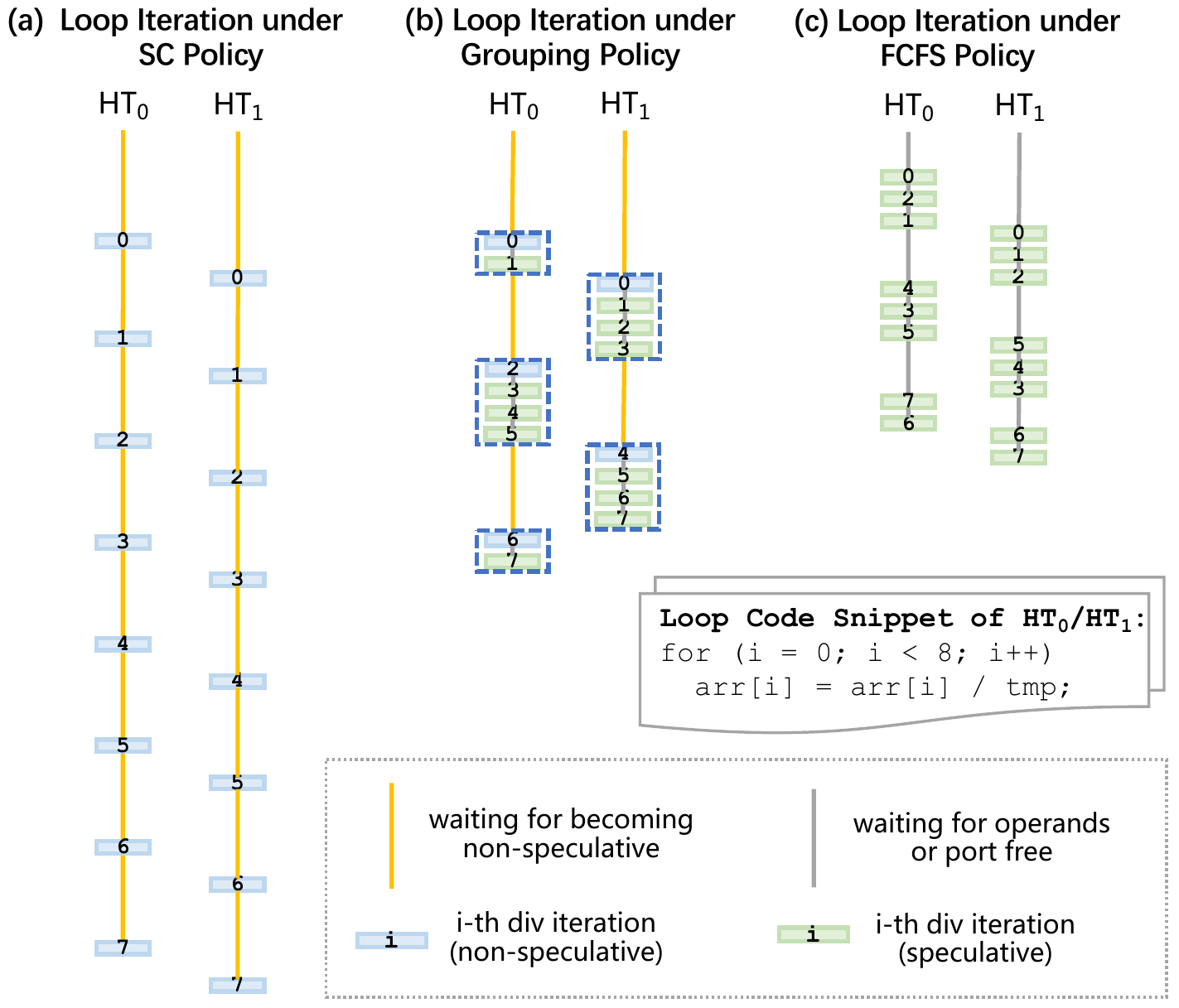}
	\caption{The timelines of loop iterations in the code snippet shown in the figure under Speculative Compression (SC), Grouping (i.e. the policy of \sys), and native FCFS policies, respectively.}
	\label{fig:example}
	\vspace{-1.5em}
\end{figure}

We first revisit the handicaps of two existing defense approaches, i.e. Time-Division-Multiplexing (TDM) and Speculation Compression (SC).
For TDM, the strict partitioning of time slices is too rigid for most workloads that may have unbalanced resource requirements.
For example, if we have two active HTs in the system, with one being more computation-intensive and the other more memory-intensive.
Using TDM, almost half of the computation and memory resources may be wasted. 
For SC, when branch instructions are issued frequently and the average branch resolution time is long (nearly 20 cycles for SPEC benchmarks according to our evaluation), the aggregated issuing delay will incur a significant performance overhead.
\autoref{fig:example}(a) shows the timelines of two HTs that execute the code snippet shown in the box.
In each iteration, the division operation within the loop body need to be delayed until the guarding branch is resolved, i.e. until the operation becomes non-speculative. If the loop iterates more than 8 times, the overhead can reach 8X as long for SC policy.

\bheading{Our Insight.} The above observation leads us to conclude that \emph{if the scheduler can batch multiple speculative operations from the same HT into a group and allow them to exclusively occupy the resource for a short period of time (as most of the speculative execution windows are relatively short), then it can avoid the delay on each group member to improve performance.} 
As shown in \autoref{fig:example}, compared to the timeline under SC policy in (a), the timeline of grouping-based policy in (b) greatly reduces the number of delays and is close to the timeline under the native FCFS policy in (c).
However, to make such grouping secure against both inter-/intra-thread SCAs, we must overcome the following two challenges: 1) We must be able to form groups within an HT, and schedule the groups from different HTs to prevent group interference from being exploited by attackers who may launch inter-thread SCAs; 2) We must be able to eliminate backward contention among speculative operations within a group to defend against intra-thread SCAs.

To address these chanllenges, we propose a priority-based scheduler, called \sys. It consists of three scheduling policies: a) Non-speculative Operations have higher Priority over speculative operations (denoted as NOP); b) Last-owner-thread's Operations have higher Priority among speculative operations from different HTs (denoted as LOP); c) Earlier Operations have higher Priority within the same HT (denoted as EOP).

Among them, the NOP and LOP policies guide the group formation and regulate inter-thread scheduling among the groups, and the EOP policy resolves the backward resource contention within a group. As shown in \autoref{fig:example}(b), the NOP policy enforce each group header operation to wait until it becomes non-speculative; LOP policy allows inner speculative operations within a group to inherit resource until another group header operation (from the other HT) become non-speculative; EOP policy inhibits the disorder of inner speculative operations within a group.

The security concept behind them is the principle of \emph{Speculative Non-Interference (SNI)}\cite{SNI}, i.e. \emph{the observable states of a speculative execution should be indistinguishable from those when the same code sequence is executed non-speculatively.} Focusing on resource contention-based covert channels, the SNI principle means that the machine states as the result of the resource allocation policies should be independent of whether the code sequence is executed speculatively or not. 
We will detail the workflow, defense mechanism and performance impact of these policies in the following subsections.

\vspace{-1.5em}
\subsection{NOP: Non-speculative Operations have higher Priority}
\begin{figure}[!t]
	\centering
	\includegraphics[width=\columnwidth]{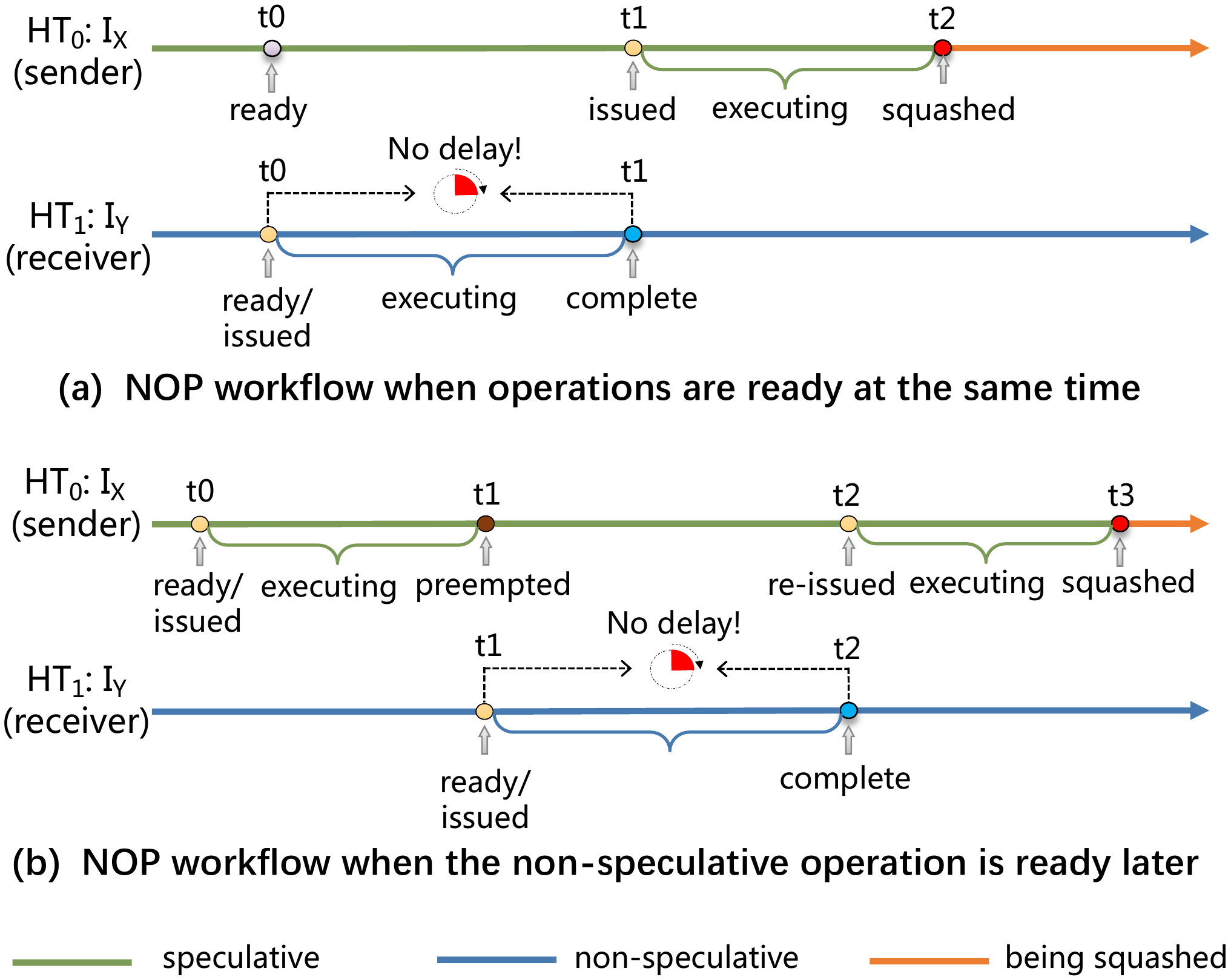}
	\caption{The workflow of the NOP policy and its defense mechanism in an inter-thread scenario. It is also the same in an intra-thread scenario.}
	\label{fig:nop}
	\vspace{-1.5em}
\end{figure}



In \sys, each non-speculative operation is treated as the header of a group. When a group header operation compete for resources together with another group inner speculative operations, it should be assigned higher and preemptive priority. 

This policy ensures that non-speculative operations from a potential receiver can access the shared resource immediately, and will not observe the interference caused by speculative operations staged by a potential sender. \autoref{fig:nop} shows such an example with two instructions from two HTs, one as a sender and the other as a receiver, competing for an issuing port. The policy takes effect in the following two scenarios.

If the port is free when the contention occurs, the non-speculative instruction will access the port immediately. As \autoref{fig:nop}(a) shows, instruction $I_X$ from $HT_0$ and instruction $I_Y$ from $HT_1$ compete for the port at $t_0$. The status of $I_X$ is speculative, and $I_Y$ is not. 
According to the NOP policy, $I_Y$ can occupy the port immediately, and $I_X$ needs to wait until the port is free. In this scenario, if $HT_0$ acts as the sender and $HT_1$ as the receiver, owing to the NOP policy on $HT_1$, no information can be transmitted.
    
If the port is currently occupied by a speculative instruction, the non-speculative instruction can preempt it immediately. As shown in \autoref{fig:nop}(b), speculative instruction $I_X$ from $HT_0$ (acting as the sender) is occupying the port exclusively at $t_0$; At $t_1$, instruction $I_Y$ (acting as the receiver) is ready and becomes non-speculative. According to the NOP policy, it can preempt the port, and $I_X$ needs to wait for being re-issued until the completion of $I_Y$ at $t_2$. In this case, the sender cannot interfere with the receiver, thus no information can be transmitted either. 


\vspace{-1em}
\subsection{LOP: Last-owner-thread's Operations have higher Priority}

If the group header operation has occupy a resource, the rest of inner speculative operations within the group can inherit the ownership without waiting for becoming non-speculative, until it is preempted by another group header operation. In other words, when multiple speculative operations from different HTs compete for a resource, \sys will give a higher priority to the competitor whose HT is the most recent non-speculative owner of the resource.

\begin{figure}[!t]
	\centering
	\includegraphics[width=\columnwidth]{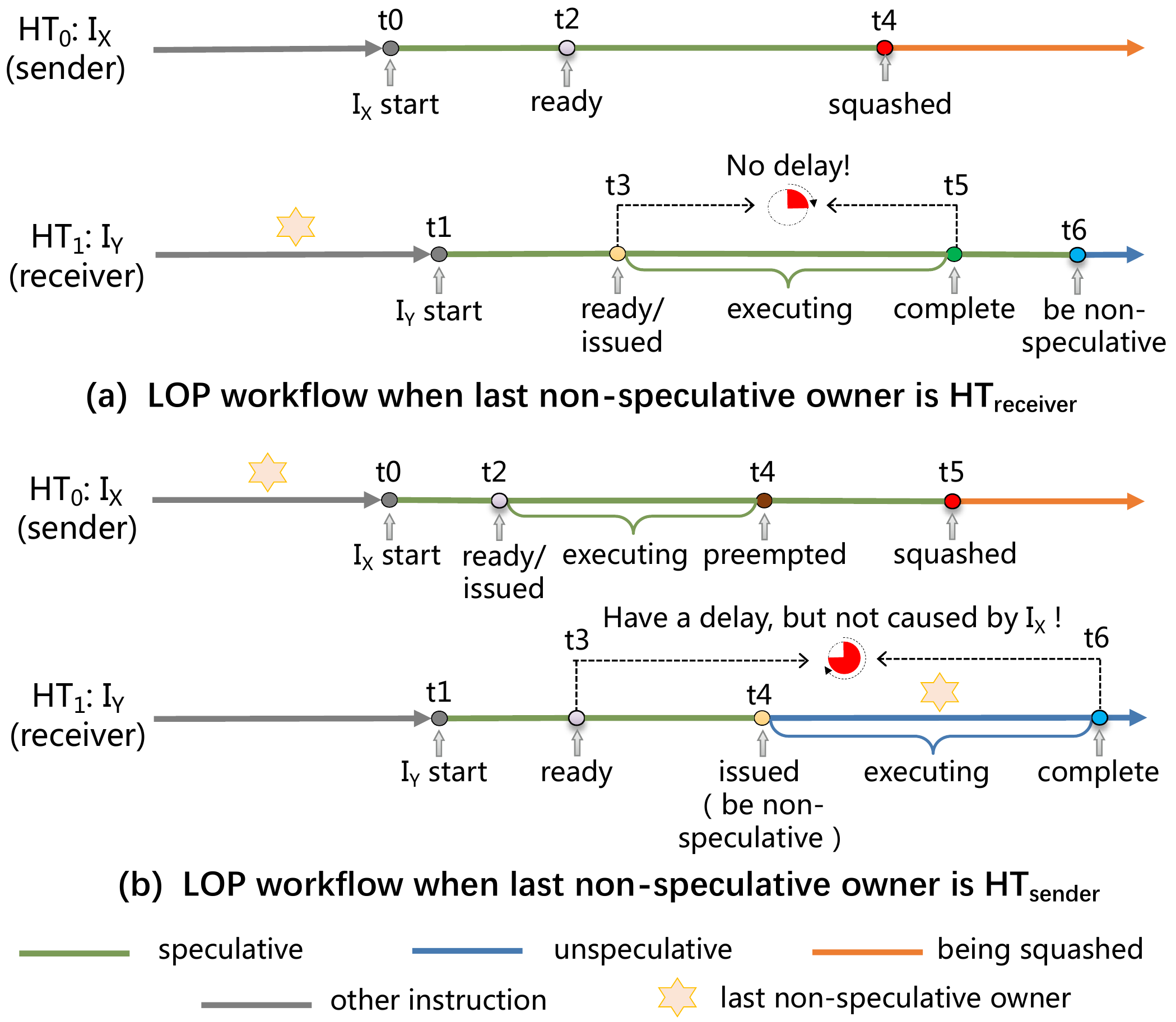}
	\caption{The workflow of the LOP policy and its defense mechanism.}
	\label{fig:lop}
	\vspace{-1.5em}
\end{figure}

Note that this policy is \emph{non-preemptive}, which means the owner HT can exclusively occupy the resource during the current period, regardless of whether it needs it or not, until the other HT preempts the resource, i.e. switches the ownership, via the NOP policy. 

Combined with the NOP policy, this policy guarantees that the resource allocation across multiple HTs is independent of any speculative operation.
It prevents the accessor of an SCA from modulating the resource to setup a covert channel.
\autoref{fig:lop} shows the workflow of the LOP policy in more details and discuss how it can defeat inter-thread SCAs.



In \autoref{fig:lop}(a), we assume $HT_1$ is a more recent non-speculative owner of the port. At $t_2$, the instruction $I_X$ of $HT_0$ is ready, but according to the LOP policy, it still cannot occupy the port even though the port is free. At $t_3$, when the instruction $I_Y$ of $HT_1$ is ready, it can be issued immediately. If $HT_0$ acts as the sender and $HT_1$  as the receiver, nothing can be transmitted because $HT_1$ observes no delay.
    
On the contrary, in \autoref{fig:lop}(b), $HT_0$ is the most recent non-speculative owner and its subsequent instruction $I_X$ (acting as the sender) can occupy the port immediately at $t_2$. Thus, the instructions $I_Y$ from $HT_1$ (acting as the receiver) needs to wait until it becomes non-speculative at $t_4$. \emph{In this case, although a delay occurs, the receiver cannot attribute it to the contention created by the sender, because no matter $I_X$ exists or not, $HT_1$ still needs to wait as it is not the most recent non-speculative owner. The delay only depends on the previous non-speculative instructions of $HT_0$, which cannot be dependent to the speculative accessor of any SCA.}

\vspace{-1em}
\subsection{EOP: Earlier Operations have higher Priority}
The above policies does not aim at intra-thread SCAs, i.e. the contention of speculative operations within the group, due to multi-issuing and out-of-order execution. Therefore, we need more information to identify potential receivers and senders and keep a receiver from observing the interference created by a sender within the same HT.

\begin{figure}[!t]
	\centering
	\includegraphics[width=\columnwidth]{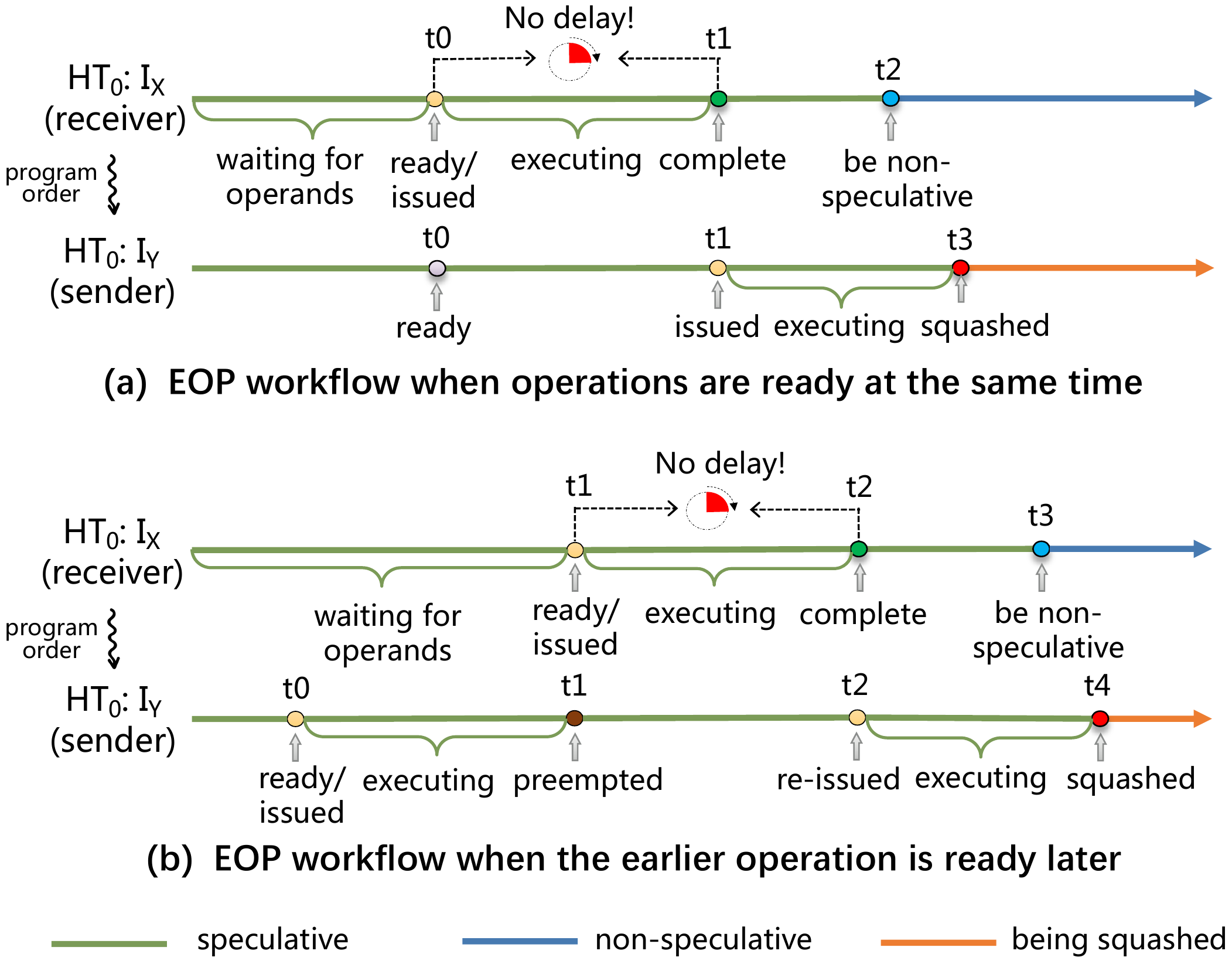}
	\caption{The workflow of the EOP policy and its defense mechanism.}
	\label{fig:eop}
	\vspace{-1.5em}
\end{figure}

With a closer look at the PoC in \autoref{fig:poc_intra}, we can notice that the receiver must be earlier than the sender in the program order. Otherwise, the completion time of the receiver will depend on the resolution time of the branch instruction at Line-8, instead of its own execution time. That is why intra-thread SCAs are called \emph{SpecRewind Attacks}~\cite{SpectreRewind}.
Based on this observation, \sys assigns a higher and preemptive priority to the earlier speculative operations than other speculative operations within the same HT.


\autoref{fig:eop} shows the EOP workflow using an example with two speculative instructions from $HT_0$. We assume the earlier one ($I_X$) serves as the receiver and the later one ($I_Y$) as the sender. There is no data dependence between them, and the issuing of receiver is delayed because its operands are not available until it encounters the sender within the scheduling window.
As shown in \autoref{fig:eop}(a), if the port is idle at $t_0$ and two instructions are both ready to be issued, the earlier receiver will be scheduled first. However, as shown in \autoref{fig:eop}(b), if the sender is issued at $t_0$ and the receiver becomes ready at $t_1$, it will immediately preempt the issuing port.

\vspace{-1em}
\subsection{The Estimation on Performance Impact}
\begin{figure}[!t]
	\centering
	\includegraphics[width=\columnwidth]{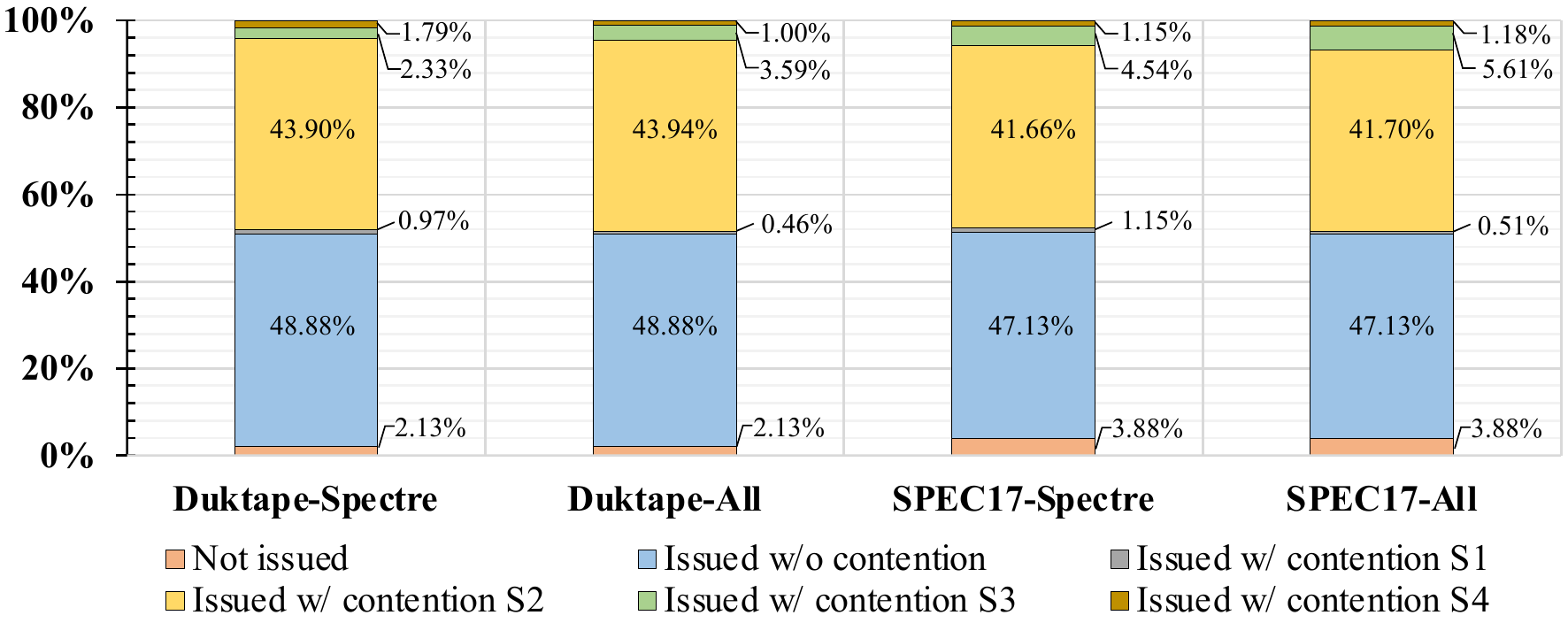}
	\caption{The distribution of various contention scenarios on a "vanilla" SMT processor. \textbf{S1}: one competitor is non-speculative and the other is speculative; \textbf{S2}: both competitors from different HTs are speculative, and the competitor from the most recent owner HT is the first to occupy the resource; \textbf{S3}: same as S2, except the competitor is not from the most recent owner HT that occupies the source first; \textbf{S4}: both competitors are speculative and from the same HT.}
	\label{fig:freq_breakup}
	\vspace{-1.5em}
\end{figure}

We collect some statistics on various contention scenarios on an insecure native SMT system to estimate the performance impact of those three policies. The configuration and the methodology of the experiments are the same as those described in \autoref{subsec:setup}, and the results are presented in \autoref{fig:freq_breakup}.

\bheading{For NOP.} The NOP policy can affect the performance in two ways. It can positively eliminate the contention caused by the wrong-path speculation. Also, it can negatively preempt the speculative execution on the correct path by some non-speculative operations, and force some of its operations to be re-executed. However, as shown in \autoref{fig:breakup_spec}, almost 50\% of instructions are not issued (3\%) or issued without contention (47\%). For these two situations, \sys will not incur any overhead. Only about 1\% of the issued instructions encounter a contention with one non-speculative and the other speculative competitor. This situation can lead to preemption under the NOP policy. Thus, we believe that the NOP policy will incur only modest overhead. 

\bheading{For LOP.} The LOP policy is designed not only to break the \emph{speculative dependence} required by all SCAs, but also to utilize the temporal locality of resource occupancy to minimize the performance impact. As shown in \autoref{fig:freq_breakup}, scenarios S2 and S3, i.e. both competitors are speculative and from different HTs, constitute 46\% of total cases. Among them, 42\% are in scenario S2, i.e. the first arriving request are from the last owner HT; While only 4\% cases are in scenario S3, i.e. the first arriving request is not from the last owner HT. The results show that the efficiency of the LOP policy used in \sys approximates to the FCFS policy used in the insecure native SMT processor.

\bheading{For EOP.} From \autoref{fig:freq_breakup}, we can see that only about 1\% of the cases are in scenario S4, i.e. both competitors are speculative and from the same HT, which may violate the EOP policy and cause preemption and re-execution with a negative impact on performance. This result indicates that the EOP policy will also have a minimal impact on the overall performance. 

Overall, the performance impact of all three scheduling policies are quite small. And subsequent performance evaluation on a simulated system, as described in \autoref{subsec:overall} and \autoref{subsec:detail_analysis}, also confirm our assertions.

\section{Key Implementation Aspects}\label{sec:impl}
\begin{figure}[!t]
    \centering
    \includegraphics[width=\columnwidth]{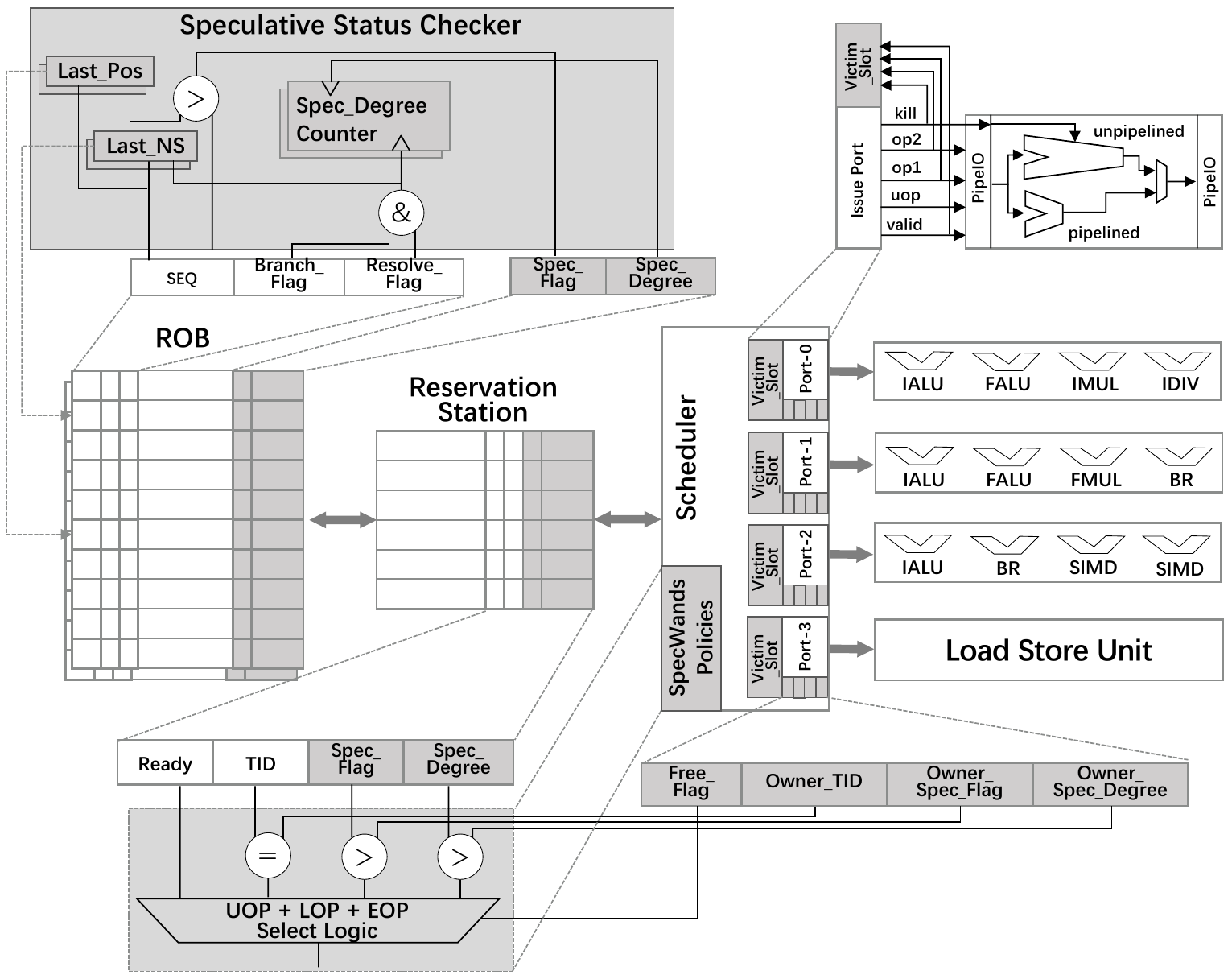}
    \caption{The framework of \sys. The shaded boxes are those introduced by \sys.}
    \label{fig:framework}
    \vspace{-1.5em}
\end{figure}

\autoref{fig:framework} presents the framework of \sys on a typical SMT microarchitecture. 
Before introducing its implementation, we need to clarify the definition on when a speculative instruction becomes "non-speculative". It can determine the defense capability and the impact on performance. Similar to earlier work such as \cite{STT, InvisiSpec, CleanupSpec}, \sys has two operating modes that correspond to two variants of definition on when an instruction has become "non-speculative".
\begin{packeditemize}
    \item \textbf{\sys-Spectre.} This operating mode only defends against SCAs that exploit vulnerabilities from branches, e.g. Spectre-PHT/BTB/RSB. Here, a speculative instruction is considered to have become \emph{non-speculative} when all its previous branch instructions have been resolved and predicted correctly. This mode has a small performance overhead because the speculative instructions from the correctly-predicted path can become non-speculative quickly and free to compete for resources as the instructions in the unsecured native processor. But its defense scope is more restrictive, and should be deployed with other mitigation schemes to form a more comprehensive defense system.
    
    \item \textbf{\sys-All.} This variant can defend against all SCAs that include existing SCAs and any future SCA. Here, an instruction is considered to be non-speculative only when it reaches the head of ROB without triggering an exception. This variant has a relatively higher performance overhead due to longer waiting time for becoming non-speculative. But it has a much wider defense scope so that it does not rely on any other mitigation schemes.
\end{packeditemize}

Based on the above definition, we can also give two variants of the definition on the instruction ordering in the EOP policy. In the \sys-ALL mode, we adhere to the conventional definition of instruction ordering, i.e. an "earlier" instruction means an "older" instruction in the original program order. In the \sys-Spectre mode, we define instruction ordering between two instructions using their relative \emph{speculative degree}, which means the number of control flow predictions (branches) exercised by the instruction. From the view of dynamic control flow graph, the instruction in the deeper basic block (dominated by more branches) has larger speculative degree. Taking the PoC in \autoref{fig:poc_intra} as an example, the division instruction at Line-13 has 2-more speculative degree than the division instruction at Line-4, since before Line-13 enters the pipeline, the speculation of previous branches at Line-8 and Line-11 must have been exercised. This variant further reduces the performance overhead of the EOP in the \sys-Spectre mode, because instructions within the same basic block have the same speculative degree, and thus no preemption and re-execution occurs.

\vspace{-1em}
\subsection{Speculative Status Checker (SSC)}
SSC plays the role to check whether an instruction is speculative or not, and compute its speculative degree if it is. Such information is needed in the NOP and EOP policies. 
SSC is associated with ROB, which provides a global program ordering and execution status of each HT. When instructions are inserted/removed in/from ROB, or ROB receives updated speculation information, SSC will be activated to check and update the speculation status of each ROB entry, and then notify the instruction scheduler. The speculation status is recorded in the following two fields in each ROB entry.
\begin{packeditemize}
    \item \textbf{Spec\_Flag}: 1-bit tag indicating whether the instruction is speculative or not.
    
    \item \textbf{Spec\_Degree}: 7-bit tag specifying the \emph{speculative degree} of the instruction.
\end{packeditemize}

In a simple implementation, SSC can scan from the header of ROB, setting the \emph{Spec\_Flag} of each instruction to non-speculative and the \emph{Spec\_Degree} to zero, until it encounters an unresolved branch. The \emph{Spec\_Flag} of the subsequent instructions are all set to speculative, and the \emph{Spec\_Degree} will be incremented with the number of scanned unsolved branches. 
However, such an implementation may incur significant overhead and power consumption when ROB is large as in modern CPUs (more than 200). In \sys, we limit the width of each scan and adopt a progressive scanning strategy. In each round, SSC only scans a fixed number of instructions (usually the same as the issue width), and records the results in three intrinsic registers for next scan. 
\begin{packeditemize}
    \item \textbf{Last\_Pos}: 8-bit field pointing to the ending entry of this scan.
    
    \item \textbf{Last\_NS}: 8-bit field pointing to the entry of the last non-speculative instruction in this scan.

    \item \textbf{Spec\_Degree\_Counter}: 8-bit field accumulating the number of unresolved branches encountered.
\end{packeditemize}

When a ROB squash occurs due to a mispeculation, SSC can quickly reset the \emph{Last\_Pos} to the position of the youngest unsquashed instruction, and recover the \emph{Last\_NS} and \emph{Spec\_Degree\_Counter} from the \emph{Spec\_Flag} and \emph{Spec\_Degree} fields of that instruction.

\vspace{-1em}
\subsection{Enhanced SMT Instruction Scheduler}
Each reservation station (RS) entries needs to be tagged with two fields: \emph{Spec\_Flag} and \emph{Spec\_Degree}, which will be updated by SSC and used by the SMT instruction scheduler. The scheduler also needs to add a \emph{Victim\_Slot} for each issuing port to temporarily store each preempted instruction. When an instruction is issued, its opcode and operands are fed into the execution unit and also stored in the \emph{Victim\_Slot}. Once the scheduler decides to preempt the instruction, the port asserts the \emph{kill} signal to the execution unit, which clears the internal state of the unpipelined unit to allow for accepting new operands on the next cycle. The \emph{kill} signal simultaneously activates the port's \emph{Victim\_Slot} to re-inserted the preempted instruction into the RS. Additionally, each port also needs an intrinsic control register to record the information for scheduling. The register includes the following four fields:

\begin{packeditemize}
    \item \textbf{Free\_Flag}: 1-bit field indicating whether the port is free or not. Only when it is cleared (i.e. free), the following four fields can be valid.
    
    \item \textbf{Owner\_TID}: 1-bit field indicating which HT is occupying the port.
    
    \item \textbf{Owner\_Spec\_Flag}: 1-bit field recording whether the occupying instruction is speculative.
    
    \item \textbf{Owner\_Spec\_Degree}: 7-bit field recording the speculative degree of the occupying instruction
\end{packeditemize}

\begin{figure}[!t]
    \centering
    \includegraphics[width=0.9\columnwidth]{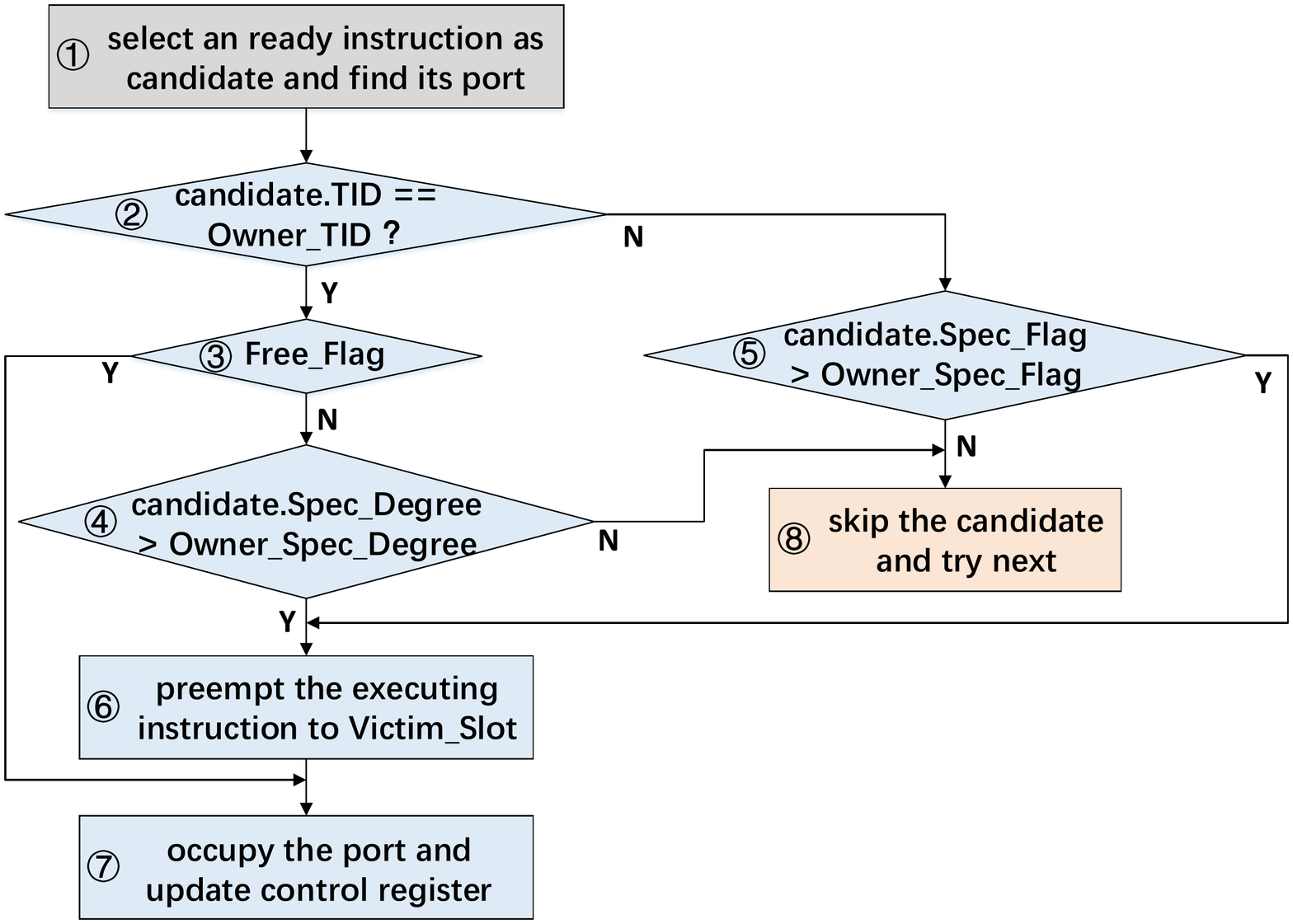}
    \caption{The workflow of the instruction issuing logic in the \sys scheduler.}
    \label{fig:workflow}
    \vspace{-1.5em}
\end{figure}

\autoref{fig:workflow} shows the workflow of the enhanced SMT instruction scheduler. 
\emph{\textbf{Step-1:}} Select a ready instruction from RS as the candidate. \emph{\textbf{Step-2:}} Determine whether the candidate's thread ID (TID) matches the port's \emph{Owner\_TID}. If they match, go to Step-3; Otherwise, go to Step-7. \emph{\textbf{Step-3:}} Check the port's \emph{Free\_Flag} to determine whether it is free or not. If it is free, go to Step-7; Otherwise, go to Step-4. \emph{\textbf{Step-4:}} If the candidate's TID matches the port's \emph{Owner\_TID}, but the port is not free, it means the port is being occupied by another instruction from the same HT. In this case, according to the EOP policy, we can just compare their speculative degrees. If the candidate wins, go to Step-6 for preemption; Otherwise, go to Step-8. \emph{\textbf{Step-5:}} If the candidate does not match the port's \emph{Owner\_TID}, it means the port is occupied by another HT. According to the NOP policy, we need to check whether the candidate is non-speculative and the owner instruction is speculative or not. If so, go to Step-6; Otherwise, go to Step-8. \emph{\textbf{Step-6:}} According to the NOP or EOP policy, preempt the occupying instruction and put it in the port's \emph{Victim\_Slot}. \emph{\textbf{Step-7:}} Issue the candidate, update the port's control register. \emph{\textbf{Step-8:}} When this step is reached, it means issuing of this candidate has failed. 
So, just skip it and schedule the next instruction.
\begin{figure*}[!t]
    \centering
    \includegraphics[width=2\columnwidth]{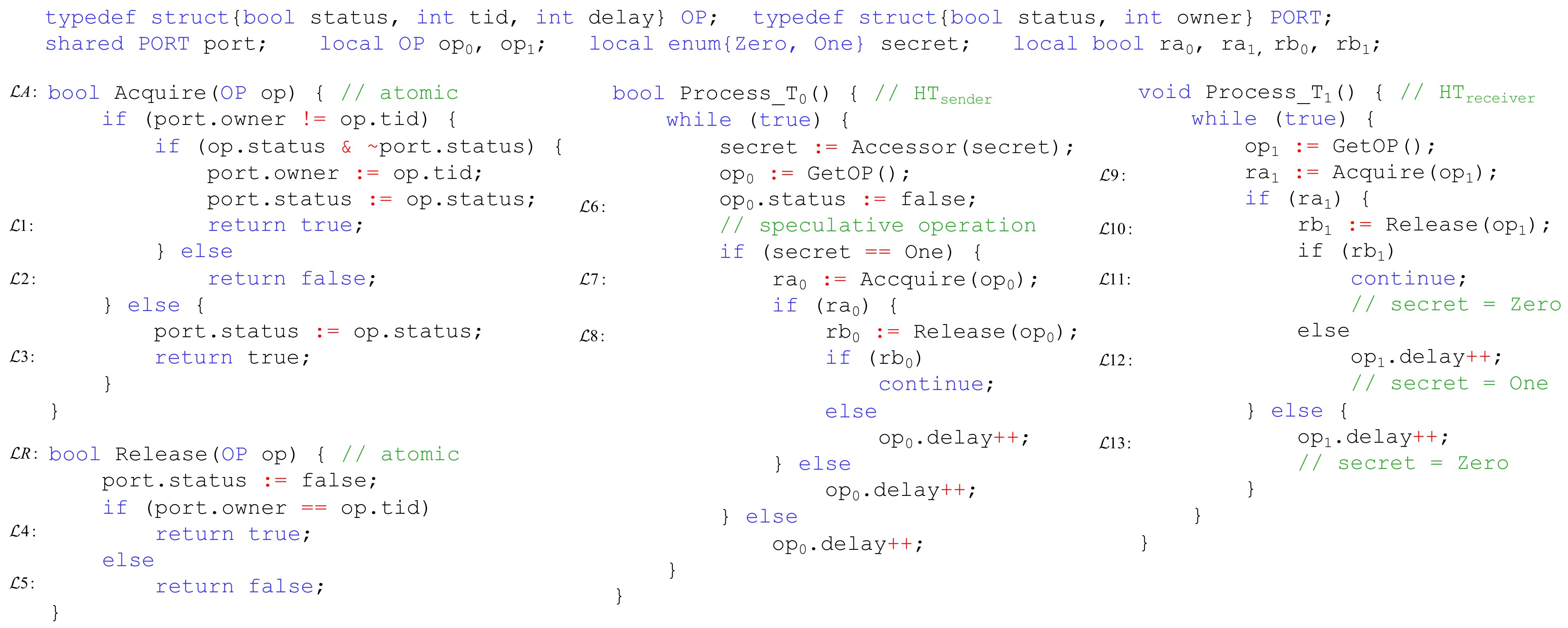}
    \caption{The pseudo code of inter-thread SCA flow for security analysis. The code is written by C-like syntax, where the assignment symbol "=" is replaced by ":=" to avoid confusion with the equality symbol "=" in the logical expression. The critical actions have been tagged with $\mathcal{L}X$, and all boolean variables' initial value are false. $Process\_T_0$ and $Process\_T_1$ are executed in parallel.}
    \label{fig:pseudo}
    \vspace{-1.5em}
\end{figure*}
\section{Security Analysis}\label{sec:analysis}
The rigorous proof of SNI principle involves formally modeling the entire workflow of scheduler, and enumerating all possible input instruction sequences. This requires significant manual effort even with the help of advanced automated proof tools. Therefore, we propose a workaround by simplifying the proof target from the attack perspective. The core proof of inter-thread SCA mitigation on 2-context SMT CPU has been completed. (full proof including intra-thread SCA mitigation is on going work). The proof is realized through Owicki-Gries method~\cite{owicki}, which is widely applied to prove the correctness of concurrent system. 

To facilitate logical expressions, we define some booleanize functions as below to map secret data and delay time to boolean value. We also define some new logical symbols, such as independent symbol "$p \indep $q" indicating $(p \notiff q) \vee (p \notiff \neg q)$, "$\otimes $p" indicating the boolean value of $p$ is unchanged after the occurance of an action, semantic definition symbol "$[\![\mathcal{A}]\!] \triangleq p$" means the state predicate $p$ after action $\mathcal{A}$ occurs, temporary property definition symbol "$\Box [\mathcal{A}_1 \cdot \mathcal{A}_2] \triangleq \mathcal{P}$" means the property $\mathcal{P}$ always been maintained after action $\mathcal{A}_1$ and $\mathcal{A}_2$ occur one after the other. 
\begin{gather}\label{eq:indicator}
    \mathcal{I}(x) = \begin{cases}
                False, & x = 0 \\
                True, & x = 1 \\
           \end{cases}
    \quad
    \mathcal{T}(x) = \begin{cases}
                False, & x = 0 \\
                True, & x > 0 \\
            \end{cases}
\end{gather}

\emph{\textbf{Step-1}}, we clarify the target property to be proved as \autoref{eq:property-0}. It implies the SNI principle that whenever attackers observe the system, the information they obtain, i.e. the delay time of the $HT_{receiver}$'s operation, is always independent to the secret data (assuming only 1-bit).
\begin{equation}\label{eq:property-0}
    \mathcal{M} \models \Box (I(secret) \indep T(op_{receiver}.delay))
\end{equation}

\emph{\textbf{Step-2}}, we abstract the attack process to two HTs, $HT_{sender}$ (tid equals zero) and $HT_{receiver}$ (tid equals one) into pseudo code (described in C-like language) in \autoref{fig:pseudo}. They compete for an issue port protected by \sys. The actions of acquiring/releasing the port, i.e. the workflow of \sys, are modularized into the function \emph{acquire()}/\emph{release()}. $HT_{receiver}$ keeps acquiring resources, while $HT_{sender}$ decides whether to compete with it based on the value of secret. There are two implications in the code: \emph{acquire()}/\emph{release()} must be executed atomically; and the port's occupying status must be false before initialization. Beyond that, we do not make any other assumptions.

\emph{\textbf{Step-3}}, we conclude the state predicates for the actions of pseudo code as shown in \autoref{fig:predicates}. To simplify the derivation, we directly compress the control flow into some critical actions, e.g. the function exits and the statements modifying the sharing variables. In the predicate, we directly refer boolean expressions and variables from C-type pseudo code. The superscript symbol "$^+$" indicates the new state of the variable after performing an action. The implicit variable "ra" indicates the return boolean value of \emph{acquire()} and "rb" indicates the return boolean value of \emph{release()}. 

\emph{\textbf{Step-4}}, according to some basic deductive rules, we can derive three invariant properties about function \emph{Accquire()}, as shown in \autoref{eq:property-1}. These properties present the conditions under which \emph{acquire()} can succeed; and the conditions under which the owner and status of the port can remain unchanged.
\begin{gather}\label{eq:property-1}
    \begin{split}
    & \Box [\mathcal{L}A] \Rightarrow \Box ([\![\mathcal{L}1]\!] \vee [\![\mathcal{L}2]\!] \vee [\![\mathcal{L}3]\!]) \Rightarrow (\mathcal{P}_1 \wedge \mathcal{P}_2 \wedge \mathcal{P}_3) \\
    & \mathcal{P}_1 \triangleq \Box (((port.owner = op.tid) \thinspace \vee \\
    & \hphantom{\mathcal{P}_1 \triangleq \Box ((} (\neg op.status \vee port.status)) \Leftrightarrow \otimes port.owner) \\
    & \mathcal{P}_2 \triangleq \Box (((port.owner \neq op.tid) \thinspace \wedge \\
    & \hphantom{\mathcal{P}_2 \triangleq \Box ((} (\neg op.status \vee port.status)) \Rightarrow \otimes port.status) \\
    & \mathcal{P}_3 \triangleq \Box (((port.owner = op.tid) \thinspace \vee \\
    & \hphantom{\mathcal{P}_3 \triangleq \Box ((} (op.status \wedge \neg port.status)) \Leftrightarrow ra)) \\
    \end{split}
    \raisetag{20pt}
\end{gather}

Similarly, we can get the invariant property about function \emph{release()}, as shown in \autoref{eq:property-2}.
\begin{gather}\label{eq:property-2}
    \begin{split}
    & \Box [\mathcal{L}R] \Rightarrow \Box ([\![\mathcal{L}4]\!] \vee [\![\mathcal{L}5]\!]) \Rightarrow \mathcal{P}_4 \qquad \qquad \qquad \qquad \qquad \\
    & \mathcal{P}_4 \triangleq \Box (((port.owner = op.tid) \Leftrightarrow rb)) \\
    & \hphantom{\mathcal{P}_4 \triangleq \Box (} \wedge \neg port.status) \\
    \end{split}
    \raisetag{20pt}
\end{gather}

\emph{\textbf{Step-5}}, we summarize the invariant properties of each single thread, which are listed in \autoref{eq:property-3}-\autoref{eq:property-5}. For $HT_{receiver}$, we consider how the result of \emph{acquire()}/\emph{release()} affects the operation delay in each loop iteration. As \autoref{eq:property-3} shows, the operation delay of $HT_{receiver}$ is absolutely dependent to the result of \emph{acquire()} and \emph{release()}.
\begin{gather}\label{eq:property-3}
    \begin{split}
    & \Box [\mathcal{L}T_1] \Rightarrow \Box ([\![\mathcal{L}11]\!] \vee [\![\mathcal{L}12]\!] \vee [\![\mathcal{L}13]\!]) \Rightarrow \mathcal{P}_5 \qquad \qquad \quad \\
    & \mathcal{P}_5 \triangleq \Box ((ra_1 \wedge rb_1) \Leftrightarrow \neg \mathcal{T}(op_{receiver}.delay))
    \end{split}
    \raisetag{20pt}
\end{gather}

While for $HT_{sender}$, we need to gradually deduce some properties of $HT_{sender}$ based on the properties obtained in \autoref{eq:property-1}, and then consider how the secret value affects the port owner and status in each loop iteration. As \autoref{eq:property-4} shows, the execution of action sequence "$\mathcal{L}6\cdot|\mathcal{L}A_0|\cdot|\!|\mathcal{L}R_0|\!|$" never change the port owner, and the port status is always false. In the equation, the symbol "$|\mathcal{A}|$" indicates that action $\mathcal{A}$ occurs conditionally and "$|\mathcal{A}|\cdot|\!|\mathcal{B}|\!|$" indicates that action $\mathcal{B}$ conditionally occurs on the premise that conditional action $\mathcal{A}$ occurs.
\begin{gather}\label{eq:property-4}
    \begin{split}
    & \Box [\mathcal{L}_6] \Rightarrow \Box ([\![\mathcal{L}6]\!]) \Rightarrow \mathcal{P}_6 \triangleq \Box (\otimes port.owner) \\
    & \Box [\mathcal{L}_6 \cdot |\mathcal{L}A_0|] \Rightarrow \Box ([\![\mathcal{L}6]\!] \wedge \mathcal{P}_1) \Rightarrow \mathcal{P}_7 \triangleq \Box (\otimes port.owner) \\
    & \Box [\mathcal{L}_6 \cdot |\mathcal{L}A_0| \cdot |\!|\mathcal{L}R_0|\!|] \Rightarrow \Box ([\![\mathcal{L}6]\!] \wedge \mathcal{P}_1 \wedge \mathcal{P}_4) \Rightarrow \mathcal{P}_8 \\
    & \mathcal{P}_8 \triangleq \Box (\otimes (port.owner \wedge \neg port.status))
    \end{split}
    \raisetag{20pt}
\end{gather}

Applying \autoref{eq:property-4} and other primary state predicts, we can reason about the independence between the secret data and port owner. In addition, we can apparently affirm that the execution of $HT_{sender}$ does not affect the operation status in $HT_{receiver}$.
\begin{gather}\label{eq:property-5}
    \begin{split}
    & \Box [\mathcal{L}T_0] \Rightarrow \Box ([\![\mathcal{L}6]\!] \vee [\![\mathcal{L}7]\!] \vee [\![\mathcal{L}8]\!]) \qquad \qquad \qquad \qquad \quad \\
    & \hphantom{\Box [\mathcal{L}T_0]} \Rightarrow \Box (\mathcal{P}6 \vee \mathcal{P}7 \vee \mathcal{P}8 \vee [\![\mathcal{L}7]\!] \vee [\![\mathcal{L}8]\!]) \Rightarrow \mathcal{P}_9 \\
    & \mathcal{P}_9 \triangleq \Box ((\mathcal{I}(secret) \indep port.owner) \wedge \\
    & \hphantom{\mathcal{P}_9 \triangleq \Box (} (\mathcal{I}(secret) \indep op_{receiver}.status)) \\
    \end{split}
    \raisetag{20pt}
\end{gather}

\begin{figure*}\label{fig:predicates}
\centering
\vspace{-1em}\quad
\subfloat{
\begin{minipage}[t]{.45\linewidth}
\begin{equation*}
\begin{split}
    [\![\mathcal{L}1]\!] \triangleq & \thinspace IF \thinspace (port.owner \neq op.tid) \thinspace THEN \\
                                    & \quad IF \thinspace (op.status \wedge \neg port.status) \thinspace THEN \\
                                    & \quad \quad \wedge (port.owner^+ = op.tid) \\
                                    & \quad \quad \wedge (port.status^+ = op.status) \wedge ra \\
    [\![\mathcal{L}3]\!] \triangleq & \thinspace IF \thinspace (port.owner = op.tid) \thinspace THEN \\
                                    & \quad \wedge (port.status^+ = op.status) \\
                                    & \quad \wedge (port.owner^+) \\
                                    & \quad \wedge ra
\end{split}
\end{equation*}
\end{minipage}
}
\quad
\subfloat{
\begin{minipage}[t]{.45\linewidth}
\begin{equation*}
\begin{split}
    [\![\mathcal{L}2]\!] \triangleq & \thinspace IF \thinspace (port.owner \neq op.tid) \thinspace THEN \\
                                    & \quad IF \thinspace (\neg op.status \vee port.status) \thinspace THEN \\
                                    & \quad \quad \wedge (ort.owner^+) \\
                                    & \quad \quad \wedge (port.status^+) \wedge \neg ra \\
    [\![\mathcal{L}4]\!] \triangleq & \thinspace IF \thinspace (port.owner = op.tid) \thinspace THEN \\
                                    & \quad \wedge \neg port.status \wedge rb \\
    [\![\mathcal{L}5]\!] \triangleq & \thinspace IF \thinspace (port.owner \neq op.tid) \thinspace THEN \\
                                    & \quad \wedge \neg port.status \wedge \neg rb
\end{split}
\end{equation*}
\end{minipage}
}
\quad
\subfloat{
\begin{minipage}[t]{.3\linewidth}
\begin{equation*}
\begin{split}
    [\![\mathcal{L}6]\!] \triangleq & \thinspace \neg op_0.status^+ \\
    [\![\mathcal{L}7]\!] \triangleq & \thinspace IF \thinspace (\mathcal{I}(secret)) \thinspace THEN \\
                                    & \quad \wedge ra_0^+ = LA_0(op) \\
    [\![\mathcal{L}11]\!] \triangleq & \thinspace IF \thinspace (ra_1) \thinspace THEN \\
                                     & \quad \thinspace IF \thinspace (rb_1) \thinspace THEN \\
                                     & \quad \quad \wedge \neg T(op_1.delay^+)
\end{split}
\end{equation*}
\end{minipage}
}
\subfloat{
\begin{minipage}[t]{.3\linewidth}
\begin{equation*}
\begin{split}
    [\![\mathcal{L}8]\!] \triangleq & \thinspace IF \thinspace (\mathcal{I}(secret)) \thinspace THEN \\
                                    & \quad \thinspace IF \thinspace (\neg ra_0) \thinspace THEN \\
                                    & \quad \quad \wedge rb_0^+ = LR_0(op) \\
    [\![\mathcal{L}12]\!] \triangleq & \thinspace IF \thinspace (ra_1) \thinspace THEN \\
                                     & \quad \thinspace IF \thinspace (\neg rb_1) \thinspace THEN \\
                                     & \quad \quad \wedge T(op_1.delay^+)
\end{split}
\end{equation*}
\end{minipage}
}
\subfloat{
\begin{minipage}[t]{.3\linewidth}
\begin{equation*}
\begin{split}
    [\![\mathcal{L}9]\!] \triangleq & \thinspace ra_1^+ = LA_1(op) \\
    [\![\mathcal{L}10]\!] \triangleq & \thinspace IF \thinspace (ra_1) \thinspace THEN \\
                                     & \quad \wedge rb_0^+ = LR_0(op) \\
    [\![\mathcal{L}13]\!] \triangleq & \thinspace IF \thinspace (\neg ra_1) \thinspace THEN \\
                                     & \quad \wedge T(op_1.delay^+)
\end{split}
\end{equation*}
\end{minipage}
}
\vspace{-0.5em}
\caption{Predicates for critical actions of the pseudo code. In pseudo code, we omit the return value of Acquire()/Release(), but in the predicates, we explicitly express their changes for verification.}\label{fig:predicates}
\vspace{-1.5em}
\end{figure*}

\emph{\textbf{Step-6}}, based on all the above properties, we can eventually prove the non-interference property between secret value and receiver operation delay. We generalize all interleaving of two parallel HTs into a sequence pattern consisting of critical actions, as shown in \autoref{eq:property-6}. The pattern implies that each interval between \emph{acquire()} and \emph{release()} in $HT_{receiver}$ can be interrupted by the actions in $HT_{sender}$, which also corresponds to the assumptions mentioned in the \autoref{sec:threat}. The \autoref{eq:property-6} indicates that only two factors, i.e. the port owner and operation status, can determine the result of \emph{acquire()}/\emph{release()} in $HT_{receiver}$. Also the inverse negation of the proposition also holds. While \autoref{eq:property-5} has prove that the secret value is independent to these two factors, we can conclude that the final property is always true.
\begin{equation}\label{eq:property-6}
    \begin{split}
    & \hat{\mathcal{L}T_0} = \mathcal{L}_6 \cdot |\mathcal{L}A_0| \cdot |\!|\mathcal{L}R_0|\!| \\
    & \mathcal{M} \models \Box [T_0 \parallel T_1] \propto \Box [(\hat{\mathcal{L}T_0})^i \cdot \mathcal{L}A_1 \cdot (\hat{\mathcal{L}T_0)}^j \cdot \mathcal{L}R_1 ] \mid \thinspace_{i, j \ge 0} \\
    & \hphantom{\mathcal{M}} \Rightarrow \Box (\mathcal{P}_1 \wedge \mathcal{P}_2 \wedge \mathcal{P}_3 \wedge (\mathcal{P}_6 \vee \mathcal{P}_7 \vee \mathcal{P}_8) \wedge \mathcal{P}_4) \\
    & \hphantom{\mathcal{M}} \Rightarrow (\mathcal{P}10 \wedge \mathcal{P}11) \Rightarrow \mathcal{P}12 \\
    & \mathcal{P}10 \triangleq \Box (((port.owner = op_{receiver}.tid) \vee op_{receiver}.status) \\
    & \hphantom{\mathcal{P}10 \triangleq \Box (} \Rightarrow (ra_1 \wedge rb_1)) \\
    & \mathcal{P}11 \triangleq \Box (((port.owner \neq op_{receiver}.tid) \wedge \neg op_{receiver}.status) \\
    & \hphantom{\mathcal{P}11 \triangleq \Box (} \Rightarrow \neg (ra_1 \wedge rb_1)) \\
    & \mathcal{P}12 \triangleq \Box (((port.owner = op_{receiver}.tid) \vee op_{receiver}.status) \\
    & \hphantom{\mathcal{P}12 \triangleq \Box (} \Leftrightarrow (ra_1 \wedge rb_1)) \\
    & (\mathcal{P}5 \wedge \mathcal{P}9 \wedge \mathcal{P}12) \Rightarrow \Box (\mathcal{I}(secret) \indep \mathcal{T}(op_{receiver}.delay))
    \end{split}
    \raisetag{20pt}
\end{equation}

\section{Evaluation}\label{sec:eval}
\subsection{Experiment Setup and Methodology}\label{subsec:setup}
We implemented a prototype of \sys on Gem5 simulator ~\cite{gem5} (version \emph{fe187de9bd}) with the O3 CPU model. The parameters of main components are shown in \autoref{tab:parameters}. We add issuing ports and configure the grouping of execution units similar to Intel Skylake microarchitecture. In performance evaluation, we firstly run SPEC CPU 2017 (rate) benchmarks with \emph{ref} input data. To cover various scheduling scenarios, we follow the methodology of previous works~\cite{SMT-COP} by selecting the benchmark pair according to their types (integer or floating-point) and program characteristics (number of branches and L2 Cache misses) as \autoref{tab:spec17} shows. To evaluate the impact on more realistic scenarios, we also run a popular embedding Javascript engine \emph{Duktape}~\cite{duktape} (version 2.6) with \emph{Sunspider}~\cite{sunspider} benchmarks as input data. \emph{SunSpider} includes the following eight categories of applications: 3D modeling, data access, bit manipulation, encryption, complex control flow, mathematical libraries, regular expression processing, and data encoding. Like other works~\cite{secSMT}, we select one program from each category with the longest execution time and adopt the tournament pairing scheme to evaluate the upper bound overhead for each program on different scheduling scenarios.

\begin{table}[!t]
  \centering
  \caption{Parameters used in the simulated micro-architecture for the baseline data. The label shown after each component indicates its sharing strategy among multiple HTs, where "\textbf{S}" means total sharing, "\textbf{P}" means fair partition, "\textbf{M}" means multiplexing with yield scheme.}
  \resizebox{\columnwidth}{!}{
    \begin{tabular}{|l|l|}
    \hline
    \hline
    \textbf{Component} & \textbf{Parameter Value} \\
    \hline
    \hline
    Core Overview  & 8-issue, out-of-order, 2-context SMT, 2Ghz \\
    \hline
    Pipeline & \tabincell{l}{ Fetcher/Decoder/Register Renamer (\textbf{M}), 64-entry RS (\textbf{P}), \\
                              256 Int / 256 FP Physical Registers (\textbf{P}), 8 Issue Ports (\textbf{S}), \\
                              92-entry ROB (\textbf{P}), 32-entry LQ (\textbf{P}), 32-entry SQ (\textbf{P})} \\
    \hline
        BPU   & Tournament branch predictor(\textbf{S}), 4096 BTB (\textbf{S}), 16 RSB (\textbf{S}) \\
    \hline
        Private L1-I Cache & \tabincell{l}{32KB 64B line, 4-way, 1 cycle RT latency, 8 MSHRs (\textbf{S})} \\
    \hline
        Private L1-D Cache & \tabincell{l}{64KB, 64B line, 8-way, 1 cycle RT latency, 8 MSHRs (\textbf{S})} \\
    \hline
        Shared L2 Cache & \tabincell{l}{2MB bank, 64B line, 16-way, 8 cycles RT local latency, \\16 cycles RT remote latency, 16 MSHRs (\textbf{S})} \\
    \hline
    \hline
    \end{tabular}%
    }
    \label{tab:parameters}
    \vspace{-1.5em}
\end{table}


\vspace{-0.5em}
\subsection{Effectiveness Evaluation}
\begin{figure}[H]
	\centering
	\includegraphics[width=\columnwidth]{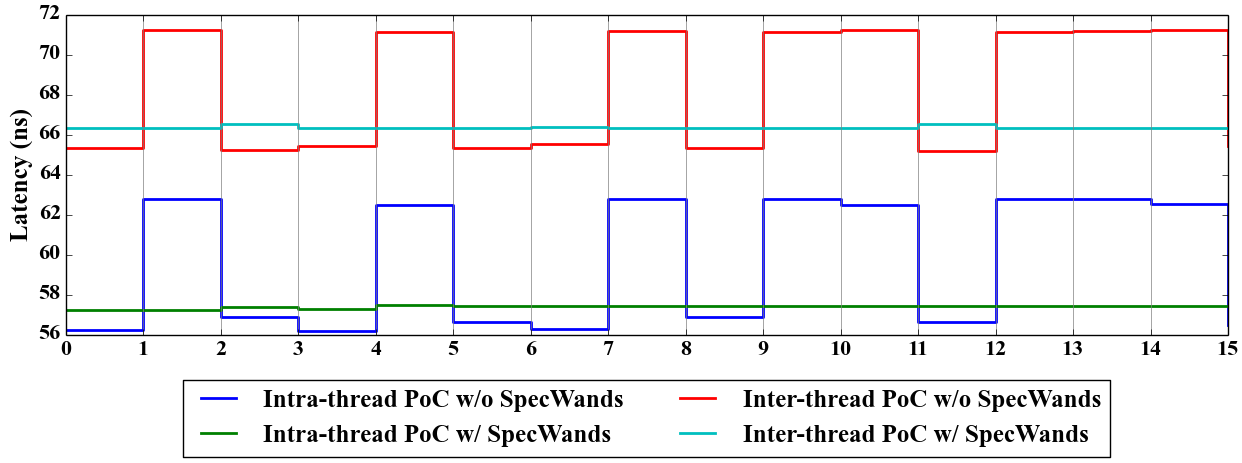}
	\caption{Latency measured by the receiver of inter/intra-thread SCA running on the native/\sys-hardened CPU. To improve legibility, we superimpose the results of inter-thread PoC above those of intra-thread PoC, and shift them upward by a constant value.}
	\label{fig:poc_evaluate}
\end{figure}

Firstly, we construct two PoCs of inter/intra-thread SCA exploiting Spectre-PHT as shown in \autoref{fig:poc_inter} and \autoref{fig:poc_intra}. To maximize the window for contention, we choose the integer division unit as the covert channel, which takes the longest time (12 cycles) to complete the computation. In PoCs, each iteration can transmit 1 bit value, repeated 100 times to reduce statistical error. For the inter-thread PoC, considering that GEM5 does not support full-system simulation in the SMT mode, and the memory space of each HT is completely isolated, it is quite challenging to synchronize the $HT_{sender}$ and $HT_{receiver}$ in each iteration. Thus, we add an instruction in the ISA dedicated to synchronize HTs on SMT, whose function is similar to \emph{pthread\_barrier\_wait()}. The simulation results of the \sys-hardened system compared to the native system are presented in \autoref{fig:poc_evaluate}. It shows that the attacker can accurately leak each bit in the native system, while unable to do so in the \sys-hardened system. 

\begin{figure}[h]
	\centering
	\includegraphics[width=\columnwidth]{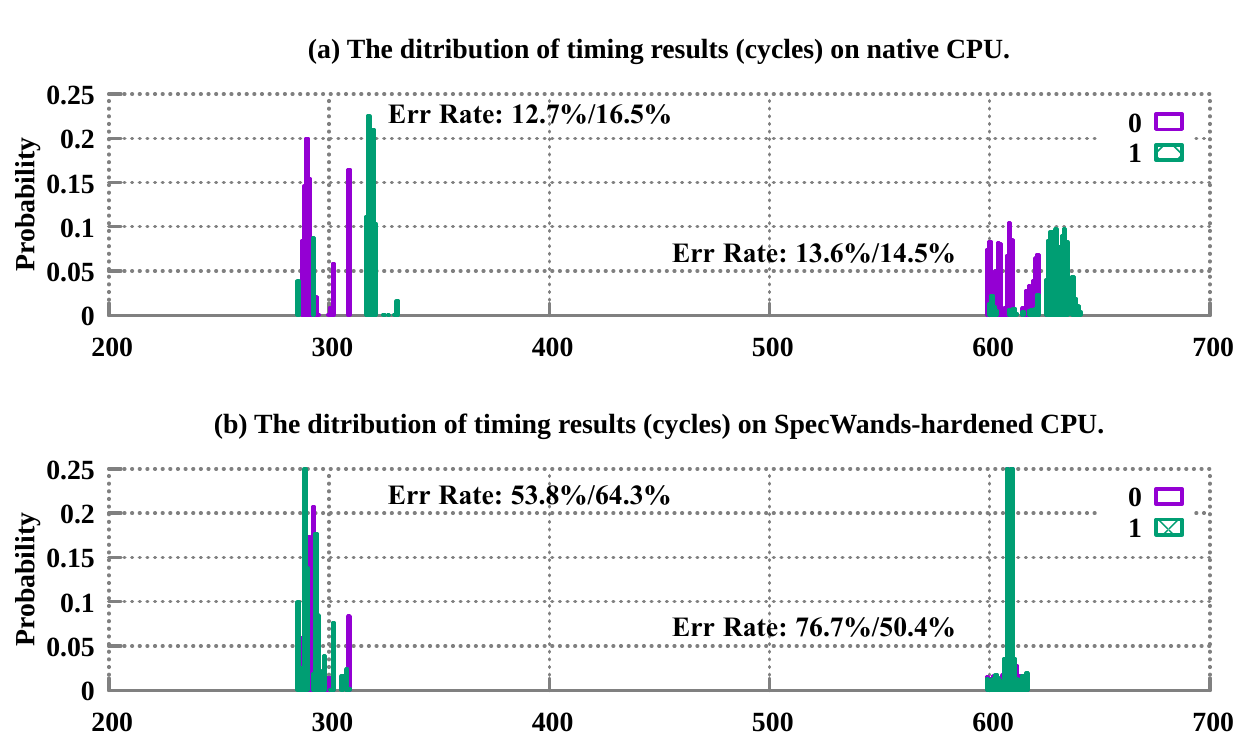}
	\caption{The distribution of latency measured by the receiver code. Similar as \autoref{fig:poc_evaluate}, we shift the result of \emph{SMoTher} attack to allow both results to be presented in a single figure. To obtain a more accurate error rate, we modify the default threshold of \emph{SMoTher} and adopt the thresholding sampling algorithm of \emph{SpectreRewind}.}
	\label{fig:attack_evaluate}
	\vspace{-1em}
\end{figure}

Second, we test \sys against two open-sourced attacks in the wild: one is \emph{SMoTher}~\cite{spectreSmother}, an inter-thread SCA that exploits Spectre-BTB vulnerability~\cite{btbTiming}; the other is \emph{SpectreRewind}~\cite{SpectreRewind}, an intra-thread SCA that exploits Meltdown vulnerability~\cite{meltdown0}. Because Meltdown vulnerability does not exists in Gem5 microarchitecture and requires full system simulation, we rewrite the process of \emph{SpectreRewind} to exploit Spectre-STL vulnerability~\cite{spectreSTL}. \autoref{fig:attack_evaluate} shows the distribution of latency and error rates for both attacks. We can see: on the native CPU, the attacker has a lower error rate to distinguish whether the transmitted bit is 1 or 0, while on the \sys-hardened CPU, the error rate is already higher than 50\%, which is equivalent to random guessing.

\begin{table}[!t]
    \centering
    \caption{SPEC 2017 benchmarks, divided into four categories based on the number of branch instructions and L2 Cache miss rate. "\textbf{I}" stands for integer programs and "\textbf{F}" stands for floating-point programs. The categorization refers to previous research~\cite{spec2017}.}
    \resizebox{\columnwidth}{!}{
    \begin{tabular}{|l|l|l|}
    \hline
    \hline
          & \textbf{Low L2 Cache Miss} & \textbf{High L2 Cache Miss} \\
    \hline
    \multirow{2}[0]{*}{\tabincell{c}{\textbf{Low} \\\textbf{BrNum}}} & \textbf{I}: x264, exchange2, perlbench & \textbf{I}: xz, xalancbmk \\
          & \textbf{F}: named, lbm, cactuBSSN & \textbf{F}: fotonik2d, bwaves \\
    \hline
    \multirow{2}[0]{*}{\tabincell{c}{\textbf{High} \\\textbf{BrNum}}} & \textbf{I}: leela, deepsjeng & \textbf{I}: gcc, omnetpp, mcf \\
          & \textbf{F}: povray, imagick, nab, parest & \textbf{F}: blender, wrf, cam4, roms \\
    \hline
    \hline
    \end{tabular}
    }
    \label{tab:spec17}
    \vspace{-1em}
\end{table}

\vspace{-1em}
\subsection{Performance Evaluation}\label{subsec:overall}
We compare the performance of \sys with two other schemes, where SMT-COP\cite{SMT-COP} represents time-division-multiplexing (TDM) scheme and STT\cite{STT} represents the speculative compression scheme. Since SMT-COP is not open-sourced, we re-implement its scheme on Gem5. 
To give a fair comparison, for SMT-COP, we do not implement other adaptive strategies that may sacrifice its security; For STT, we lift the protection for the persistent-channel components, such as cache/TLB and PHT/BTB/RSB. 
Similar to \sys, STT also has two defense modes, i.e. Spectre-Mode and All-Mode (called Futristic-Mode in their paper).

\begin{figure*}[!t]
	\centering
	\includegraphics[width=2\columnwidth]{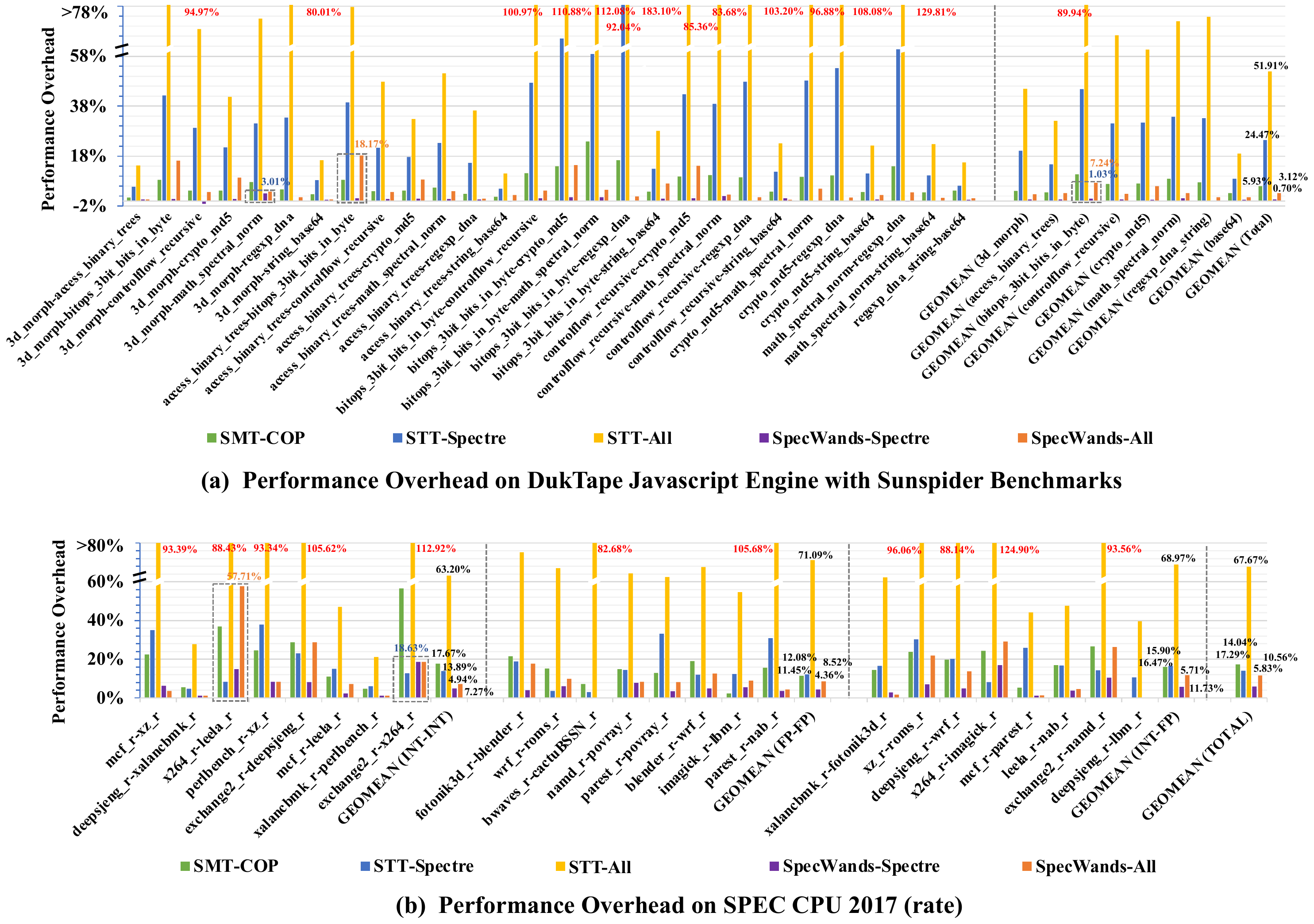}
	\caption{The performance overhead of SMT-COP, STT-Spectre, STT-All, \sys-Spectre and \sys-All on Duktape Javascript engine and SPEC CPU 2017 (rate) benchmarks. The bars taller than 80\% are truncated, whose maximum values are marked in red.}
	\label{fig:performance_spec}
	\vspace{-1.5em}
\end{figure*}

\autoref{fig:performance_spec} (a) and (b) show the performance overhead of the three defenses for Duktape Javascript engine and SPEC CPU 2017 (rate), respectively. From the figure, we can see that the overall performance overhead of \sys-Spectre/-All are 0.70\%/3.12\% and 5.83\%/10.56\%, which are much lower than 24.47\%/51.91\% and 14.04\%/67.47\% of STT, and also much lower than 5.93\% and 17.29\% of SMT-COP. For each benchmark pair, \sys significantly outperforms STT in both defense modes. And this advantage also shows in comparison between \sys and SMT-COP, except for a few cases such as \emph{povray-calculix} and \emph{x264\_r-leela\_r} in SPEC 2017. To facilitate more detail analysis, we record other statistics such as the latency of operand-ready instructions waiting for issuing, the busy rate of issuing port, etc., as shown in \autoref{tab:stats}.

\begin{table}[!t]
    \centering
    \caption{Mean statistic of running Duktape/SPEC on each system. "\textbf{1}":SMT-COP, "\textbf{2}":STT-Spectre, "\textbf{3}":STT-All, "\textbf{4}":\sys-Spectre, "\textbf{5}":\sys-All, "\textbf{-Sp}": Spectre-Mode. "\textbf{LW}": Latency of operand-ready instruction waiting for issuing; "\textbf{LT}": Latency of issue-Trying, which equals the latency between the moment of instruction being issued first time and the moment being issued successfully (last time); "\textbf{LX}": Latency of instruction execution. "\textbf{BR}": Busy rate of issuing ports; "\textbf{UR}": Utilization rate of issuing width. "\textbf{FR}": Full-event occurrence rate of reservation station. }
    \resizebox{\columnwidth}{!}{
    \begin{tabular}{|l|l|l|l|l|l|l|}
    \hline
    \hline
    & \textbf{LW (c)} & \textbf{LT (c)} & \textbf{LX (c)} & \textbf{BR (\%)} & \textbf{UR (\%)} & \textbf{FR (\%)} \\
    \hline
    \hline
    \textbf{1} & 10.8/14.8 & 0.6/0.8 & 2.6/3.9 & 16.1/24.1 & 42.9/51.4 & 10.5/20.5 \\
    \hline
    \textbf{2} & 13.0/19.1 & 0.2/0.3 & 2.5/3.7 & 19.6/22.8 & 85.7/91.4 & 21.2/18.4 \\
    \hline
    \textbf{3} & 22.3/33.3 & 0.2/0.2 & 2.5/3.5 & 18.8/19.9 & 90.5/94.6 & 31.6/37.7 \\
    \hline
    \textbf{4} & 7.58/11.5 & 0.2/0.5 & 2.5/3.8 & 20.5/29.1 & 86.3/82.7 & 6.23/14.2 \\
    \hline
    \textbf{5} & 10.5/13.2 & 0.4/0.4 & 2.5/3.6 & 19.4/22.6 & 74.6/77.8 & 9.81/18.2 \\
    \hline
    \hline
    \end{tabular}
    }
    \label{tab:stats}
    \vspace{-1.5em}
\end{table}

\bheading{Versus STT.} The main overhead of STT comes from the need to wait for the dependant instructions to become non-speculative, which may cause the pipeline to stall. \autoref{tab:stats} indeed shows the system hardened by STT has a much larger issuing waiting latency than the other two (almost by 20 to 30 cycles).
The trend is even more pronounced when the program has a larger branch resolution time or a dependency chain with higher L2 Cache miss, such as the pairs containing \emph{control\_flow\_recursive}, \emph{math\_spectral\_norm}, \emph{crypto\_md5} in Duktape engine, and \emph{perlbench\_r}, \emph{mcf\_r}, \emph{parest\_r} in SPEC 2017. But, once the instructions are allowed to be issued in STT, HTs are free to compete for the resources as in an unprotected system.
So, its utilization of issuing bandwidth is the most efficient among the three.

\bheading{Versus SMT-COP.} The main overhead of SMT-COP is the time waiting for an HT's own time slice, which often leads to longer issuing waiting latency and much lower issuing port utilization (below 50\%) as \autoref{tab:stats} shows. 
For memory intensive programs, such as \emph{bitops\_3bit\_bits\_in\_byt}, \emph{exp\_dna\_string} in Duktape engine, and \emph{x264\_r}, \emph{wrf\_r}, and \emph{exchange2\_r} in SPEC 2017, such low utilization rates become more serious. The advantage of SMT-COP is its simplicity to implement, but its scalability is the worst among three schemes. The length of the time slice depends on the longest completion time of any unpipelined execution unit, and the waiting time is proportionate to the number of HTs supported on the CPU. 

Additionally, a common factor contributing to the performance overhead of STT and SMT-COP is longer issue delay, which frequently stalls the pipeline and drags down the overall performance. Compared to STT and SMT-COP, \sys incurs a main overhead in the form of issuing waiting latency induced by the LOP policy, as well as preemption/re-execution overhead induced by the NOP and EOP policies. Nevertheless, its overall overhead remains relatively low. The highest overheads observed in \sys-Spectre/-All are 18.63\%/57.71\%, respectively, which occur in the \emph{exchange2\_r-x264\_r} and \emph{x264\_r-leela\_r} workloads of SPEC 2017. This still falls short of the highest overhead observed in STT (37.81\% and 124.90\%) and is comparable to the highest overhead in SMT-COP (56.59\%). Moreover, as illustrated in \autoref{fig:performance_spec} (b) for realistic applications, the upper bound impact of \sys-Spectre/-All in different scenarios are approximately 1\%/7\%, with worst case scenarios not exceeding 3\%/15\%. These values are far more acceptable to developers compared to the extreme cases observed in the other two defenses.


\begin{figure}[!t]
	\centering
	\includegraphics[width=\columnwidth]{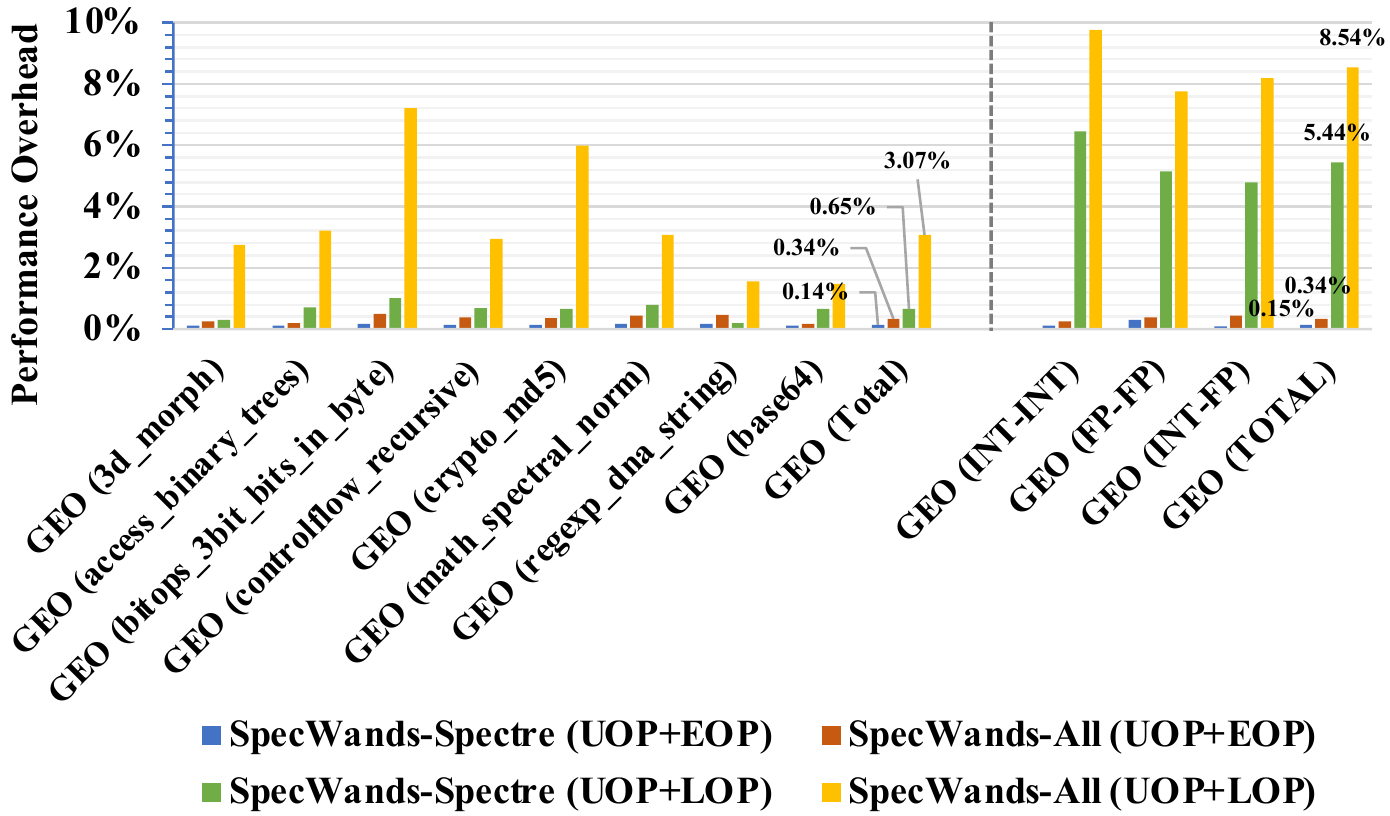}
	\caption{The performance overhead of \sys using different combinations of policies on Duktape engine and SPEC 2017.}
	\label{fig:breakup_spec}
	\vspace{-1.5em}
\end{figure}

\vspace{-1em}
\subsection{More Detailed Analysis on Performance Overhead}\label{subsec:detail_analysis}
First, we analyze the impact of different scheduling policies on the overall performance. Based on their defense scopes, we separate the policies into two groups: 1) enabling NOP and EOP policies only for intra-thread SCAs; 2) enabling NOP and LOP policies only for inter-thread SCAs.
The results are shown in \autoref{fig:breakup_spec}. The performance overhead of the NOP+EOP is around 0.1\% to 0.3\%. This is because most of the issuing port contention comes from speculative instructions. Thus, the preemption caused by NOP occurs only very infrequently. And since the native scheduler tends to issue older ones when it confronts multiple ready instructions within a HT, the preemption caused by EOP also occurs infrequently. 
In contrast, the overhead of the NOP+LOP is relatively high at around 3\%-8\%, which basically constitutes most of the overall overhead. 
It shows that, although we try to exploit the temporal locality of the issuing port using LOP, it still cannot satisfy the bandwidth demand of all HTs.

\begin{figure}[!t]
	\centering
	\includegraphics[width=\columnwidth]{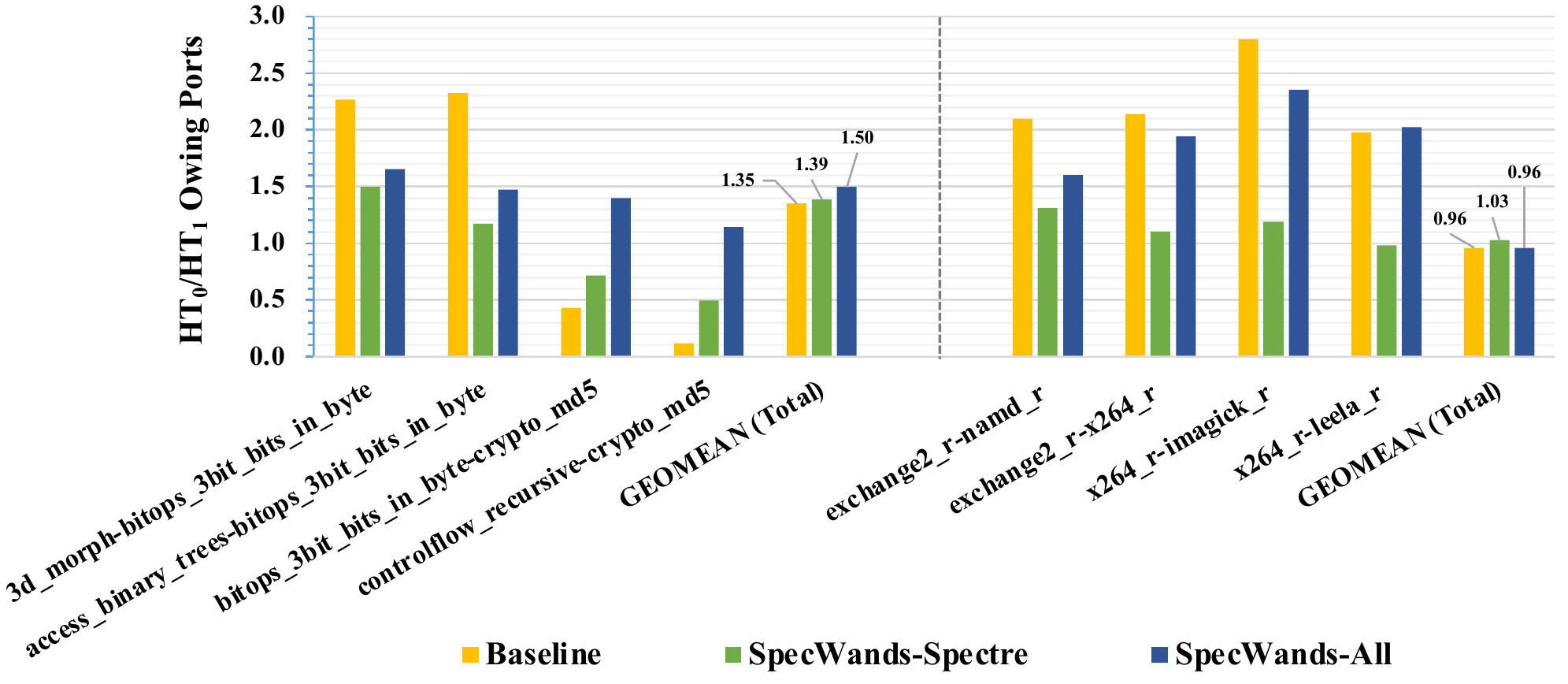}
	\caption{The ratios of issue ports owned by two HTs ($HT_0$/$HT_1$). Under the LOP policy, the number of issue ports owned by each HT are sampled per 10,000 cycles, and then their arithmetic average is calculated.}
	\label{fig:owing_ratio}
	\vspace{-1.5em}
\end{figure}

Next, we analyze the benchmark pairs that have a high performance overhead under the NOP+LOP policies. We sample the number of issuing ports occupied by each HT and calculate the ratio of that number for $HT_0$ and $HT_1$ under native FCFS policy (as baseline) and \sys, respectively. The results are shown in \autoref{fig:owing_ratio}. We find that, for those program pairs that have a high overhead under LOP, one of them must be a dominant program that has a higher occupancy rate under the native policy, such as \emph{3bit\_bits\_in\_byte}, \emph{exchange2\_r} and \emph{x264\_r}. When using the LOP policy, the non-dominant HT can have a larger share of the issuing port, which can slow down the dominant HT as a result. For this situation, one workaround is to referring more non-speculative access history of each HT for resource allocation instead of direct ownership inheriting. Such revised policy may give more opportunities to the HT that exactly needs the resource more eagerly. We will further evaluate more improvement solutions in the future works. 

\vspace{-1em}
\subsection{Power Consumption Evaluation}
We applied McPAT\cite{mcpat} (version 1.3) to model the power consumption of SMT-COP, STT, and \sys. The results are shown in \autoref{tab:power}. 
We only modify the scheduler instead of introducing new RAM components, thus the incurred hardware cost and static power consumption (such as gate leakage, sub-threshold leakage) are negligible. 
Here we only measure dynamic power when running Duktape engine and SPEC 2017.
To ensure a more accurate measurement, we patch the code of \emph{Speculative Status Checker} and \emph{Enhanced Scheduler} in \sys, as well as the code of \emph{Data Flow Tracking} and \emph{Tainting/Untainting} in STT, so that these actions can be reflected in the statistics of relevant pipeline components, i.e. execution units (EUs), reservation station (RS), and ROB.

As we can see from \autoref{tab:power}, \sys has a slightly higher power overhead on Execution Units compared to SMT-COP and STT, which mainly comes from the preemption and the re-execution required in the NOP and EOP policies. STT also consumes more power than \sys on RS and ROB accesses because both \emph{Taint} and \emph{Untaint} operations in STT require frequent accesses to these two components.

\begin{table}[!h]
    \centering
    \caption{The overhead of dynamic power consumption running Duktape/SPEC on each system. Because the running time and executed instructions of each benchmarks pairs differ under different systems, we calculate the average power consumption per nanosecond.}
    \resizebox{\columnwidth}{!}{
    \begin{tabular}{|l|l|l|l|l|l|}
    \hline
    \hline
    & \textbf{EUs (\%)} & \textbf{RS (\%)} & \textbf{ROB (\%)} \\
    \hline
    \hline
    \textbf{SMT-COP} & 1.99/2.50 & 2.01/4.32 & 1.85/2.78 \\
    \hline
    \textbf{STT-Sp} & 1.14/2.89 & 7.96/9.12 & 10.34/12.03 \\
    \hline
    \textbf{STT-All} & 2.09/2.99 & 9.97/12.92 & 14.63/17.21 \\
    \hline
    \textbf{\sys-Sp} & 1.90/5.01 & 2.86/6.75 & 6.16/10.09 \\
    \hline
    \textbf{\sys-All} & 2.63/6.59 & 4.17/9.26 & 6.81/10.92 \\
    \hline
    \hline
    \end{tabular}
    }
    \label{tab:power}
    \vspace{-1.5em}
\end{table}

\section{Discussion}
In addition to issuing ports and execution units, recent research~\cite{secSMT} has shown that other micro-architectural components that can incur inter-thread contention, such as Instruction Fetcher, Decoder, Reservation Stations, and Load Queue, also require similar attention. Here, we discuss the extension of \sys to those components.

\bheading{Instruction Fetcher \& Decoder.} These components are usually shared among multiple HTs using a round-robin policy. However, when one HT is stalled due to an I-cache miss or pipeline stall, its time slice will be borrowed by the other HT, which can affect its execution time. Such optimization can thus be exploited as a covert channel. Specifically, fetching/decoding operations can be divided into two categories: \emph{prediction-based} and \emph{resolving-based}. Prediction-based operations are controlled by the prediction units. The attacker can encode them by manipulating the prediction history via speculative branch instructions. To block this path, we can just prohibit speculative branch instructions from polluting the prediction units until they are committed. Resolving-based operations are triggered when the execution of corresponding branch instructions completes and the mispredictions are detected. If these branch instructions are also dominated by other earlier speculative branches, these operations can also be modulated to transmit the secret. For this situation, we can apply the NOP+LOP policies of \sys to determine whether the time slice borrowing will be approved. The speculative status of operation-based fetching/decoding operations can also be checked with its corresponding branch instructions by the SSC module as \autoref{sec:impl} describe.

\bheading{Reservation Station (LS) \& Load Queue (LQ).}
In some aggressive CPU, both RS and LQ adopt a constrained sharing strategy that sets a threshold of occupying entry number for each HT (usually larger than half). If one HT occupies most of entries and does not release them (achieved by manufacturing multiple cache misses), the number of available entries for the other HT will be suppressed, and thus can be exploited as a covert channel. To harden these components, we can adopt a hybrid strategy where half of the entries are set aside for fair partition between two HTs to avoid starvation, and then apply the NOP+LOP policies to the remaining half of the entries to maintain the performance benefit of temporal locality.
\section{Related Work}\label{sec:relwk}
In addition to \emph{speculation compression} and \emph{time-division-multiplexing} (TDM), there are other approaches that can mitigate SCAs. Depending on their blocking schemes, we classify them into the following three categories.

\bheading{Prediction Manipulation Prevention.} This type of mitigations only targets Spectre-type SCAs. They work by isolating/flushing/bypassing the control prediction units, which prevent them from being trained by attackers to launch illegal speculative accesses. When Spectre-V1/V2 attacks were disclosed, Intel added the \emph{Indirect Branch Control Mitigation} features (including IBRS/STIBP/IBPB) to its new processors~\cite{IBRS-STIBP-IBPB}. It prevents the states of prediction units from being transferred between user/kernel modes, non-root/root modes, and across different HTs. For Spectre-V3 attacks, Intel recommended inserting an \emph{lfense} instruction or enable \emph{Speculative Store Bypass Disable} (SSBD) mitigation before a critical \emph{store} instruction~\cite{SSBD}, which is similar to the ARM's recommendation~\cite{SSBB}. 
Some software mitigations are also proposed. For example, \emph{SLH}~\cite{meltdown1} tries to replace branch instructions with \emph{cmov} instructions, and \emph{Retpoline}~\cite{retpoline} converts all indirect \emph{jmp} instructions to \emph{call-ret} gadgets, which can bypass poisoned PHT and BTB. Compare with \sys, the above schemes can defend against SCAs for certain types of Spectre vulnerabilities only. Also, due to the lack of prediction accuracy, they will incur a higher performance overhead than \sys.

\bheading{Domain Access Isolation.} This type of mitigations is proposed for Meltdown-type TEAs. The core idea of their defense is to enforce the isolation between different domains, which prevents the secret data from being accessed by illegal cross-domain speculation. As a typical example, \emph{KPTI}~\cite{KPTI} and \emph{USC}~\cite{USC} unmap the critical entries of kernel page table from the user page table, which can foil a speculative access from the user space to the kernel space during its address translation. \emph{SF-Xen}~\cite{sfxen} introduces a similar scheme to isolate the hypervisor from untrusted guest VMs. Meanwhile, Chrome and Webkit browsers~\cite{SiteIsolation} leverage index masking and pointer poisoning to eliminate the overlap between the address spaces of different websites. 
At the hardware level, \emph{OISA}~\cite{OISA} extends the RISC-V ISA to add dedicated memory access instructions for the secret domain, which ensure normal memory access cannot touch the secret domain speculatively. \emph{ConTExT}~\cite{ConTExT} modifies speculative data forwarding of cross-domain accesses to fill nonsense values for subsequent dependant instructions. In contrast to \sys, these schemes all require the assistance of upper-layer software, and can only protect data in a specified domain against illegal speculative accesses from other domains. They cannot alleviate various SCAs based on Spectre-like vulnerabilities within the domain.

\bheading{Resource Contention Elimination.} The original intent of such approaches is to defeat traditional side channel attacks based on resource contention. In some scenarios, they can also be used to defend against SCAs. Since a native X86 processor does not support turning on/off SMT dynamically at runtime, \emph{DDM}\cite{DDM} emulates this function through the \emph{HLT} instruction and abstracts it into a user-friendly OS interface. 
\emph{Partial-SMT}\cite{partialSMT} first redesigns the user-level thread library so that users can define different thread groups according to their trust level, then it leverages affinity-scheduling features in the Linux kernel to make sure that different groups are scheduled to different physical cores. 
\emph{Hyperspace}\cite{hyperspace} modifies the compiler for Intel SGX development to instrument an enclave program and allows it to launch a shadow thread located on the same physical core. It thus prevents the contention from being tampered by an attacker. 
Same as \emph{domain access isolation}, these methods still need software support to specify the protected threads. But from the perspective of TEAs, any thread or gadget can be compromised as the sender of a covert channel. It may not be practical to manually specify all of the protected objects. 
In addition, these methods only consider inter-thread contention and cannot be applied to intra-thread contention.

\section{Conclusion}\label{sec:conclusion}
In this paper, we propose a priority-based scheduler, called \sys, to defend against \emph{transient execution attacks} (TEAs) that exploit inter-/intra-thread contention on a system component as a covert channel. \sys contains three scheduling policies: a) Non-speculative Operations have higher Priority (NOP). It allows non-speculative operations to preempt speculative operations at any time; b) Last-owner-thread's Operations have higher Priority (LOP). It allocates the resource to the speculative operations which belong to the thread occupying the resource most recently. c) Earlier Operations have higher Priority (EOP). It gives the earlier speculative operations in a thread higher and preemptive priority over the later speculative operation within the same thread. These three policies batch multiple continuous speculative operations into a group, which can exclusively occupy the resource for a certain period of time without any delay. The performance evaluation shows that \sys has a significant performance advantage over other state-of-the-art approaches such as speculative compression and time-division-multiplexing.


\bibliographystyle{IEEEtran}
\bibliography{refs}

\begin{thebibliography}{10}
\providecommand{\url}[1]{#1}
\csname url@samestyle\endcsname
\providecommand{\newblock}{\relax}
\providecommand{\bibinfo}[2]{#2}
\providecommand{\BIBentrySTDinterwordspacing}{\spaceskip=0pt\relax}
\providecommand{\BIBentryALTinterwordstretchfactor}{4}
\providecommand{\BIBentryALTinterwordspacing}{\spaceskip=\fontdimen2\font plus
\BIBentryALTinterwordstretchfactor\fontdimen3\font minus
  \fontdimen4\font\relax}
\providecommand{\BIBforeignlanguage}[2]{{%
\expandafter\ifx\csname l@#1\endcsname\relax
\typeout{** WARNING: IEEEtran.bst: No hyphenation pattern has been}%
\typeout{** loaded for the language `#1'. Using the pattern for}%
\typeout{** the default language instead.}%
\else
\language=\csname l@#1\endcsname
\fi
#2}}
\providecommand{\BIBdecl}{\relax}
\BIBdecl

\bibitem{fpTiming}
W.-M. Hu, ``Lattice scheduling and covert channels,'' in \emph{Proceedings 1992
  IEEE Computer Society Symposium on Research in Security and Privacy}.\hskip
  1em plus 0.5em minus 0.4em\relax IEEE Computer Society, 1992, pp. 52--52.

\bibitem{SIMDTiming}
M.~Andrysco, D.~Kohlbrenner, K.~Mowery, R.~Jhala, S.~Lerner, and H.~Shacham,
  ``On subnormal floating point and abnormal timing,'' in \emph{2015 IEEE
  Symposium on Security and Privacy}.\hskip 1em plus 0.5em minus 0.4em\relax
  IEEE, 2015, pp. 623--639.

\bibitem{portTiming}
A.~C. Aldaya, B.~B. Brumley, S.~ul~Hassan, C.~P. Garc{\'\i}a, and N.~Tuveri,
  ``Port contention for fun and profit,'' in \emph{2019 IEEE Symposium on
  Security and Privacy (SP)}.\hskip 1em plus 0.5em minus 0.4em\relax IEEE,
  2019, pp. 870--887.

\bibitem{TEAEvaluation}
C.~Canella, J.~Van~Bulck, M.~Schwarz, M.~Lipp, B.~Von~Berg, P.~Ortner,
  F.~Piessens, D.~Evtyushkin, and D.~Gruss, ``A systematic evaluation of
  transient execution attacks and defenses,'' in \emph{28th USENIX Security
  Symposium}, 2019, pp. 249--266.

\bibitem{foreshadow}
J.~Van~Bulck, M.~Minkin, O.~Weisse, D.~Genkin, B.~Kasikci, F.~Piessens,
  M.~Silberstein, T.~F. Wenisch, Y.~Yarom, and R.~Strackx, ``Foreshadow:
  Extracting the keys to the intel $\{$SGX$\}$ kingdom with transient
  out-of-order execution,'' in \emph{27th USENIX Security Symposium}, 2018, pp.
  991--1008.

\bibitem{mds1}
S.~Van~Schaik, A.~Milburn, S.~{\"O}sterlund, P.~Frigo, G.~Maisuradze,
  K.~Razavi, H.~Bos, and C.~Giuffrida, ``Ridl: Rogue in-flight data load,'' in
  \emph{2019 IEEE Symposium on Security and Privacy (SP)}, 2019, pp. 88--105.

\bibitem{LVI}
J.~Van~Bulck, D.~Moghimi, M.~Schwarz, M.~Lippi, M.~Minkin, D.~Genkin, Y.~Yarom,
  B.~Sunar, D.~Gruss, and F.~Piessens, ``Lvi: Hijacking transient execution
  through microarchitectural load value injection,'' in \emph{2020 IEEE
  Symposium on Security and Privacy (SP)}.\hskip 1em plus 0.5em minus
  0.4em\relax IEEE, 2020, pp. 54--72.

\bibitem{spectreSTL}
M.~Schwarz, C.~Canella, L.~Giner, and D.~Gruss, ``Store-to-leak forwarding:
  Leaking data on meltdown-resistant cpus (updated and extended version),''
  \emph{arXiv preprint arXiv:1905.05725}, 2019.

\bibitem{swapgs}
A.~LUȚAȘ and D.~LUȚAȘ, ``Bypassing kpti using the speculative behavior of
  the swapgs instruction,
  https://i.blackhat.com/eu-19/thursday/eu-19-lutas-bypassing-kpti-using-the-speculative-behavior-of-the-swapgs-instruction-wp.pdf,''
  in \emph{2019 BlackHat Europe Conference}, 2019.

\bibitem{netspectre}
M.~Schwarz, M.~Schwarzl, M.~Lipp, J.~Masters, and D.~Gruss, ``Netspectre: Read
  arbitrary memory over network,'' in \emph{European Symposium on Research in
  Computer Security}, 2019, pp. 279--299.

\bibitem{spectreInterference}
M.~Behnia, P.~Sahu, R.~Paccagnella, J.~Yu, Z.~Zhao, X.~Zou, T.~Unterluggauer,
  J.~Torrellas, C.~Rozas, A.~Morrison \emph{et~al.}, ``Speculative interference
  attacks: Breaking invisible speculation schemes,'' \emph{arXiv preprint
  arXiv:2007.11818}, 2020.

\bibitem{meltdown0}
M.~Lipp, M.~Schwarz, D.~Gruss, T.~Prescher, W.~Haas, A.~Fogh, J.~Horn,
  S.~Mangard, P.~Kocher, D.~Genkin \emph{et~al.}, ``Meltdown: Reading kernel
  memory from user space,'' in \emph{27th USENIX Security Symposium}, 2018, pp.
  973--990.

\bibitem{spectre0}
P.~Kocher, J.~Horn, A.~Fogh, , D.~Genkin, D.~Gruss, W.~Haas, M.~Hamburg,
  M.~Lipp, S.~Mangard, T.~Prescher, M.~Schwarz, and Y.~Yarom, ``Spectre
  attacks: Exploiting speculative execution,'' in \emph{40th IEEE Symposium on
  Security and Privacy (S\&P'19)}, 2019.

\bibitem{spectreSmother}
A.~Bhattacharyya, A.~Sandulescu, M.~Neugschwandtner, A.~Sorniotti, B.~Falsafi,
  M.~Payer, and A.~Kurmus, ``Smotherspectre: exploiting speculative execution
  through port contention,'' in \emph{Proceedings of the 2019 ACM SIGSAC
  Conference on Computer and Communications Security}, 2019, pp. 785--800.

\bibitem{STT}
J.~Yu, M.~Yan, A.~Khyzha, A.~Morrison, J.~Torrellas, and C.~W. Fletcher,
  ``Speculative taint tracking (stt): A comprehensive protection for
  speculatively accessed data,'' in \emph{Proceedings of the 52nd Annual
  IEEE/ACM International Symposium on Microarchitecture}, 2019.

\bibitem{NDA}
O.~Weisse, I.~Neal, K.~Loughlin, T.~F. Wenisch, and B.~Kasikci, ``Nda:
  Preventing speculative execution attacks at their source,'' in
  \emph{Proceedings of the 52nd Annual IEEE/ACM International Symposium on
  Microarchitecture}, 2019, pp. 572--586.

\bibitem{SpecShield}
K.~Barber, A.~Bacha, L.~Zhou, Y.~Zhang, and R.~Teodorescu, ``Specshield:
  Shielding speculative data from microarchitectural covert channels,'' in
  \emph{2019 28th International Conference on Parallel Architectures and
  Compilation Techniques (PACT)}.\hskip 1em plus 0.5em minus 0.4em\relax IEEE,
  2019, pp. 151--164.

\bibitem{SMT-COP}
D.~Townley and D.~Ponomarev, ``Smt-cop: Defeating side-channel attacks on
  execution units in smt processors,'' in \emph{2019 28th International
  Conference on Parallel Architectures and Compilation Techniques (PACT)},
  2019, pp. 43--54.

\bibitem{SMT-COPImproved}
U.~Nezir, B.~Lus, and G.~Kucuk, ``Improved resource scheduling for lightweight
  smt-cop,'' in \emph{2021 6th International Conference on Computer Science and
  Engineering (UBMK)}.\hskip 1em plus 0.5em minus 0.4em\relax IEEE, 2021, pp.
  575--580.

\bibitem{secSMT}
M.~Taram, X.~Ren, A.~Venkat, and D.~Tullsen, ``Secsmt: Securing smt processors
  against contention-based covert channels,'' in \emph{USENIX Security
  Symposium}, 2022.

\bibitem{SpectreRewind}
J.~Fustos, M.~Bechtel, and H.~Yun, ``Spectrerewind: Leaking secrets to past
  instructions,'' in \emph{Proceedings of the 4th ACM Workshop on Attacks and
  Solutions in Hardware Security}, 2020, pp. 117--126.

\bibitem{woSMT}
T.~Rokicki, C.~Maurice, and M.~Schwarz, ``Cpu port contention without smt,'' in
  \emph{European Symposium on Research in Computer Security}.\hskip 1em plus
  0.5em minus 0.4em\relax Springer, 2022, pp. 209--228.

\bibitem{SNI}
M.~Guarnieri, B.~K{\"o}pf, J.~F. Morales, J.~Reineke, and A.~S{\'a}nchez,
  ``Spectector: Principled detection of speculative information flows,'' in
  \emph{2020 IEEE Symposium on Security and Privacy (SP)}.\hskip 1em plus 0.5em
  minus 0.4em\relax IEEE, 2020, pp. 1--19.

\bibitem{smt0}
D.~M. Tullsen, S.~J. Eggers, and H.~M. Levy, ``Simultaneous multithreading:
  Maximizing on-chip parallelism,'' in \emph{Proceedings of the 22nd annual
  international symposium on Computer architecture}, 1995, pp. 392--403.

\bibitem{channelSurvey}
Q.~Ge, Y.~Yarom, D.~Cock, and G.~Heiser, ``A survey of microarchitectural
  timing attacks and countermeasures on contemporary hardware,'' \emph{J.
  Cryptogr. Eng.}, vol.~8, no.~1, pp. 1--27, 2018.

\bibitem{TEASurvey}
W.~Xiong and J.~Szefer, ``Survey of transient execution attacks,'' \emph{arXiv
  preprint arXiv:2005.13435}, 2020.

\bibitem{KPart}
N.~El-Sayed, A.~Mukkara, P.-A. Tsai, H.~Kasture, X.~Ma, and D.~Sanchez,
  ``Kpart: A hybrid cache partitioning-sharing technique for commodity
  multicores,'' in \emph{2018 IEEE International Symposium on High Performance
  Computer Architecture (HPCA)}, 2018, pp. 104--117.

\bibitem{DAWG}
V.~Kiriansky, I.~Lebedev, S.~Amarasinghe, S.~Devadas, and J.~Emer, ``Dawg: A
  defense against cache timing attacks in speculative execution processors,''
  in \emph{2018 51st Annual IEEE/ACM International Symposium on
  Microarchitecture (MICRO)}, 2018, pp. 974--987.

\bibitem{RPcache}
Z.~Wang and R.~B. Lee, ``New cache designs for thwarting software cache-based
  side channel attacks,'' pp. 494--505, 2007.

\bibitem{scatterCache}
M.~Werner, T.~Unterluggauer, L.~Giner, M.~Schwarz, D.~Gruss, and S.~Mangard,
  ``$\{$ScatterCache$\}$: Thwarting cache attacks via cache set
  randomization,'' in \emph{28th USENIX Security Symposium (USENIX Security
  19)}, 2019, pp. 675--692.

\bibitem{RTDetect}
J.~Cho, T.~Kim, S.~Kim, M.~Im, T.~Kim, and Y.~Shin, ``Real-time detection for
  cache side channel attack using performance counter monitor,'' \emph{Applied
  Sciences}, vol.~10, no.~3, p. 984, 2020.

\bibitem{MLDetect}
J.~Depoix and P.~Altmeyer, ``Detecting spectre attacks by identifying cache
  side-channel attacks using machine learning,'' \emph{Advanced Microkernel
  Operating Systems}, vol.~75, 2018.

\bibitem{InvisiSpec}
M.~Yan, J.~Choi, D.~Skarlatos, A.~Morrison, C.~Fletcher, and J.~Torrellas,
  ``Invisispec: Making speculative execution invisible in the cache
  hierarchy,'' in \emph{2018 51st Annual IEEE/ACM International Symposium on
  Microarchitecture (MICRO)}.\hskip 1em plus 0.5em minus 0.4em\relax IEEE,
  2018, pp. 428--441.

\bibitem{CleanupSpec}
G.~Saileshwar and M.~K. Qureshi, ``Cleanupspec: An" undo" approach to safe
  speculation,'' in \emph{Proceedings of the 52nd Annual IEEE/ACM International
  Symposium on Microarchitecture}, 2019, pp. 73--86.

\bibitem{Muontrap}
S.~Ainsworth and T.~M. Jones, ``Muontrap: Preventing cross-domain spectre-like
  attacks by capturing speculative state,'' in \emph{2020 ACM/IEEE 47th Annual
  International Symposium on Computer Architecture (ISCA)}.\hskip 1em plus
  0.5em minus 0.4em\relax IEEE, 2020, pp. 132--144.

\bibitem{SpecBox}
B.~Tang, C.~Wu, Z.~Wang, L.~Jia, P.-C. Yew, Y.~Cheng, Y.~Zhang, C.~Wang, and
  G.~Xu, ``Specbox: A label-based transparent speculation scheme against
  transient execution attacks,'' \emph{IEEE Transactions on Dependable and
  Secure Computing}, 2022.

\bibitem{DDM}
Y.~Zhang, Z.~Zhu, and D.~Meng, ``Ddm: A demand-based dynamic mitigation for smt
  transient channels,'' in \emph{2019 IEEE Intl Conf on Parallel \& Distributed
  Processing with Applications, Big Data \& Cloud Computing, Sustainable
  Computing \& Communications, Social Computing \& Networking
  (ISPA/BDCloud/SocialCom/SustainCom)}.\hskip 1em plus 0.5em minus 0.4em\relax
  IEEE, 2019, pp. 614--621.

\bibitem{partialSMT}
X.~Wu, Y.~He, Q.~Zhou, H.~Ma, L.~He, W.~Wang, and L.~Chen, ``Partial-smt:
  Core-scheduling protection against smt contention-based attacks,'' in
  \emph{2020 IEEE 19th International Conference on Trust, Security and Privacy
  in Computing and Communications (TrustCom)}.\hskip 1em plus 0.5em minus
  0.4em\relax IEEE, 2020, pp. 378--385.

\bibitem{contentionAware}
S.~Blagodurov, S.~Zhuravlev, and A.~Fedorova, ``Contention-aware scheduling on
  multicore systems,'' \emph{ACM Transactions on Computer Systems (TOCS)},
  vol.~28, no.~4, pp. 1--45, 2010.

\bibitem{underDome}
M.~Escouteloup, R.~Lashermes, J.~Fournier, and J.-L. Lanet, ``Under the dome:
  preventing hardware timing information leakage,'' in \emph{International
  Conference on Smart Card Research and Advanced Applications}.\hskip 1em plus
  0.5em minus 0.4em\relax Springer, 2021, pp. 233--253.

\bibitem{spectreLD}
\BIBentryALTinterwordspacing
Intel. (2021) Speculative load disordering / cve-2021-33149. [Online].
  Available:
  \url{https://www.intel.com/content/www/us/en/developer/articles/technical/software-security-guidance/advisory-guidance/speculative-load-disordering.html}
\BIBentrySTDinterwordspacing

\bibitem{RSRR}
\BIBentryALTinterwordspacing
------. (2018) Rogue system register read / cve-2018-3640 / intel-sa-00115.
  [Online]. Available:
  \url{https://www.intel.com/content/www/us/en/developer/articles/technical/software-security-guidance/advisory-guidance/rogue-system-register-read.html}
\BIBentrySTDinterwordspacing

\bibitem{owicki}
S.~Owicki and D.~Gries, ``Verifying properties of parallel programs: An
  axiomatic approach,'' \emph{Communications of the ACM}, vol.~19, no.~5, pp.
  279--285, 1976.

\bibitem{gem5}
N.~Binkert, B.~Beckmann, G.~Black, S.~K. Reinhardt, A.~Saidi, A.~Basu,
  J.~Hestness, D.~R. Hower, T.~Krishna, S.~Sardashti \emph{et~al.}, ``The gem5
  simulator,'' \emph{ACM SIGARCH computer architecture news}, vol.~39, no.~2,
  pp. 1--7, 2011.

\bibitem{duktape}
\BIBentryALTinterwordspacing
(2020) Duktape javascript engine. [Online]. Available:
  \url{https://duktape.org}
\BIBentrySTDinterwordspacing

\bibitem{sunspider}
\BIBentryALTinterwordspacing
Webkit. (2020) Sunspider javascript benchmarks (1.0). [Online]. Available:
  \url{https://webkit.org/perf/sunspider/sunspider.html}
\BIBentrySTDinterwordspacing

\bibitem{btbTiming}
O.~Ac{\i}i{\c{c}}mez, {\c{C}}.~K. Ko{\c{c}}, and J.-P. Seifert, ``Predicting
  secret keys via branch prediction,'' in \emph{Cryptographers’ Track at the
  RSA Conference}.\hskip 1em plus 0.5em minus 0.4em\relax Springer, 2007, pp.
  225--242.

\bibitem{spec2017}
A.~Limaye and T.~Adegbija, ``A workload characterization of the spec cpu2017
  benchmark suite,'' in \emph{2018 IEEE International Symposium on Performance
  Analysis of Systems and Software (ISPASS)}.\hskip 1em plus 0.5em minus
  0.4em\relax IEEE, 2018, pp. 149--158.

\bibitem{mcpat}
S.~Li, J.~H. Ahn, R.~D. Strong, J.~B. Brockman, D.~M. Tullsen, and N.~P.
  Jouppi, ``Mcpat: An integrated power, area, and timing modeling framework for
  multicore and manycore architectures,'' in \emph{Proceedings of the 42nd
  annual ieee/acm international symposium on microarchitecture}, 2009, pp.
  469--480.

\bibitem{IBRS-STIBP-IBPB}
\BIBentryALTinterwordspacing
Intel, ``Speculative execution side channel mitigations,'' pp. 2--7, 2018.
  [Online]. Available:
  \url{https://www.intel.com/content/dam/develop/external/us/en/documents/336996-speculative-execution-side-channel-mitigations.pdf}
\BIBentrySTDinterwordspacing

\bibitem{SSBD}
\BIBentryALTinterwordspacing
------, ``Speculative execution side channel mitigations,'' pp. 10--11, 2018.
  [Online]. Available:
  \url{https://www.intel.com/content/dam/develop/external/us/en/documents/336996-speculative-execution-side-channel-mitigations.pdf}
\BIBentrySTDinterwordspacing

\bibitem{SSBB}
\BIBentryALTinterwordspacing
ARM, ``Arm® instruction set architecture: for armv8-a architecture,'' 2022.
  [Online]. Available:
  \url{https://developer.arm.com/docs/ddi0597/h/base-instructions-alphabetic-order/ssbb-speculative-store-bypass-barrier}
\BIBentrySTDinterwordspacing

\bibitem{meltdown1}
V.~Kiriansky and C.~Waldspurger, ``Speculative buffer overflows: Attacks and
  defenses,'' \emph{arXiv preprint arXiv:1807.03757}, 2018.

\bibitem{retpoline}
M.~F.~A. Kadir, J.~K. Wong, F.~Ab~Wahab, A.~F. A.~A. Bharun, M.~A. Mohamed, and
  A.~H. Zakaria, ``Retpoline technique for mitigating spectre attack,'' in
  \emph{2019 6th International Conference on Electrical and Electronics
  Engineering (ICEEE)}, 2019, pp. 96--101.

\bibitem{KPTI}
Linux, ``The current state of kernel page-table isolation,
  https://lwn.net/articles/741878/.''

\bibitem{USC}
J.~Behrens, A.~Cao, C.~Skeggs, A.~Belay, M.~F. Kaashoek, and N.~Zeldovich,
  ``Efficiently mitigating transient execution attacks using the unmapped
  speculation contract,'' in \emph{14th USENIX Symposium on Operating Systems
  Design and Implementation (OSDI 20)}, 2020, pp. 1139--1154.

\bibitem{sfxen}
H.~Xia, D.~Zhang, W.~Liu, I.~Haller, B.~Sherwin, and D.~Chisnall, ``A
  secret-free hypervisor: Rethinking isolation in the age of speculative
  vulnerabilities,'' in \emph{2022 IEEE Symposium on Security and Privacy
  (SP)}.\hskip 1em plus 0.5em minus 0.4em\relax IEEE Computer Society, 2022,
  pp. 1544--1544.

\bibitem{SiteIsolation}
C.~Reis, A.~Moshchuk, and N.~Oskov, ``Site isolation: Process separation for
  web sites within the browser,'' in \emph{28th USENIX Security Symposium},
  2019, pp. 1661--1678.

\bibitem{OISA}
J.~Yu, L.~Hsiung, M.~El'Hajj, and C.~W. Fletcher, ``Data oblivious isa
  extensions for side channel-resistant and high performance computing,'' in
  \emph{The Network and Distributed System Security Symposium (NDSS)}, 2019.

\bibitem{ConTExT}
M.~Schwarz, M.~Lipp, C.~Canella, R.~Schilling, F.~Kargl, and D.~Gruss,
  ``Context: A generic approach for mitigating spectre.'' in \emph{NDSS}, 2020.

\bibitem{hyperspace}
G.~Chen, W.~Wang, T.~Chen, S.~Chen, Y.~Zhang, X.~Wang, T.-H. Lai, and D.~Lin,
  ``Racing in hyperspace: Closing hyper-threading side channels on sgx with
  contrived data races,'' in \emph{2018 IEEE Symposium on Security and Privacy
  (SP)}.\hskip 1em plus 0.5em minus 0.4em\relax IEEE, 2018, pp. 178--194.

\end{thebibliography}


\begin{IEEEbiography}[{\includegraphics[width=0.9in,height=1.2in,clip,keepaspectratio]{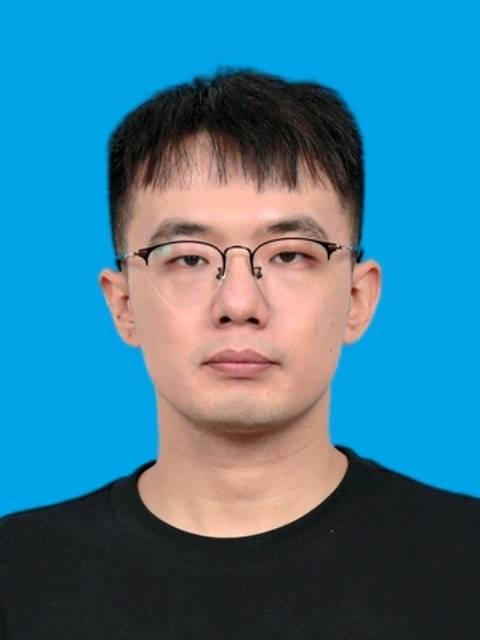}}]{Bowen Tang} is currently working toward the PhD degree in the Institute of Computing Technology, Chinese Academy of Sciences. His research interests include system security, bug detection and virtualization.
\end{IEEEbiography}

\vspace{-3 em}

\begin{IEEEbiography}[{\includegraphics[width=0.9in,height=1.2in,clip,keepaspectratio]{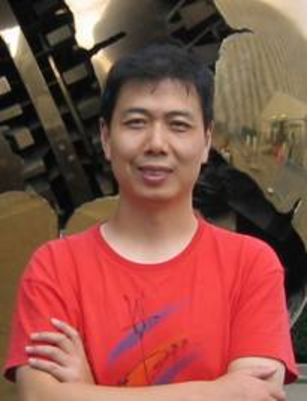}}]{Chenggang Wu} is a professor at Institute of Computing Technology, Chinese Academy of Sciences. He has served on the program committees of many major conferences. His research interests include the dynamic compilation, virtualization, bug detection on concurrent program, and system security.
\end{IEEEbiography}

\vspace{-3 em}

\begin{IEEEbiography}[{\includegraphics[width=0.9in,height=1.2in,clip,keepaspectratio]{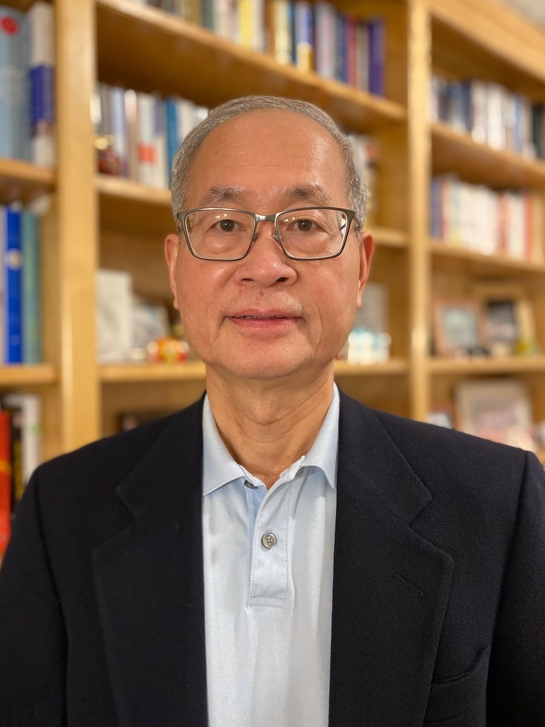}}]{Pen-Chung Yew} is a professor in CSE Department, University of Minnesota, and was the head of the department and the holder of the William-Norris Land-Grant chair professor between 2000 and 2005. His current research interests include system virtualization, compilers and architectural issues related multi-core/many-core systems. He is a IEEE fellow.
\end{IEEEbiography}

\vspace{-3 em}

\begin{IEEEbiography}[{\includegraphics[width=0.9in,height=1.2in,clip,keepaspectratio]{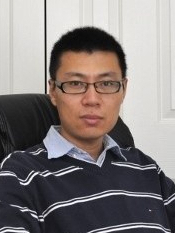}}]{Yinqian Zhang} is a professor of CSE Department, Southern University of Science and Technology (SUSTech). Before joining SUSTech in 2021, he was an associate professor at CSE Department of Ohio State University. His research interest is computer system security, with particular emphasis on cloud computing security, OS security and side-channel security.
\end{IEEEbiography}

\vspace{-3 em}

\begin{IEEEbiography}[{\includegraphics[width=0.9in,height=1.2in,clip,keepaspectratio]{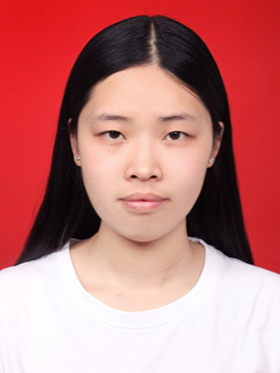}}]{Mengyao Xie} received her PhD degree in the Institute of Computing Technology, Chinese Academy of Sciences in 2022, and then has been working there until now. Her research interests include system security and virtualization.
\end{IEEEbiography}

\vspace{-3 em}

\begin{IEEEbiography}[{\includegraphics[width=0.9in,height=1.2in,clip,keepaspectratio]{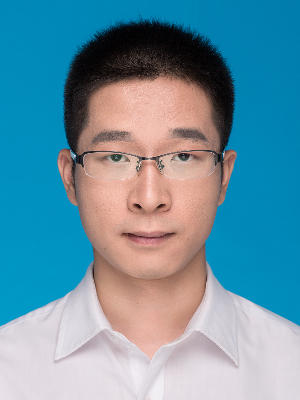}}]{Yuanming Lai} received the M.S. degree at Huazhong University of Science and Technology in 2016. Now he is in Institute of Computing Technology, Chinese Academy of Sciences. His research interests include system security and machine learning.
\end{IEEEbiography}

\vspace{-3 em}

\begin{IEEEbiography}[{\includegraphics[width=0.9in,height=1.2in,clip,keepaspectratio]{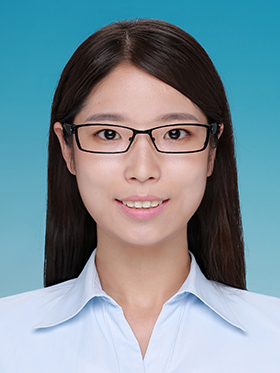}}]{Yan Kang} received the M.S. degree at Beijing University of Aeronautics and Astronautics (BUAA) in 2017, Now she is working in Institute of Computing Technology, Chinese Academy of Sciences. Her research interests include software and system security.
\end{IEEEbiography}

\vspace{-3 em}

\begin{IEEEbiography}[{\includegraphics[width=0.9in,height=1.2in,clip,keepaspectratio]{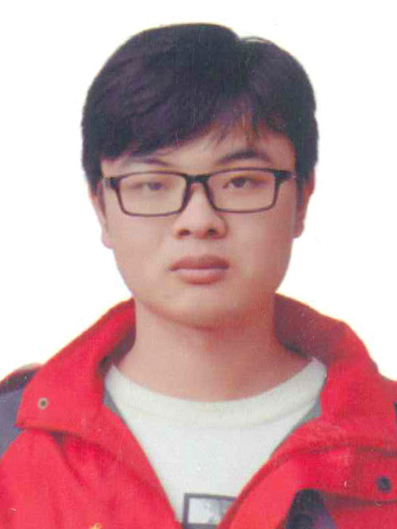}}]{Wei Wang} received his M.S. degree at Capital Normal University in 2021. He is currently working in the Institute of Computing Technology, Chinese Academy of Sciences, Beijing, China. His research interests include software security, adversarial attack and robustness.
\end{IEEEbiography}

\vspace{-3 em}

\begin{IEEEbiography}[{\includegraphics[width=0.9in,height=1.2in,clip,keepaspectratio]{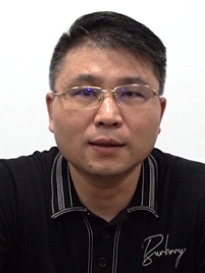}}]{Qiang Wei} is a professor at State Key Laboratory of Mathematical Engineering and Advanced Computing of China. His main research interests are network and information system security, including software vulnerability analysis, cloud computing security, etc.
\end{IEEEbiography}

\vspace{-3 em}

\begin{IEEEbiography}[{\includegraphics[width=0.9in,height=1.2in,clip,keepaspectratio]{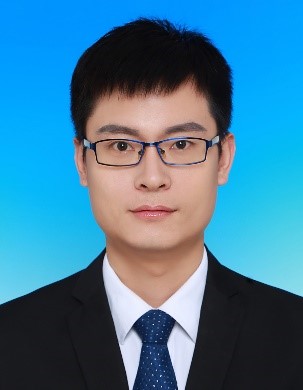}}]{Zhe Wang} is an associate professor at Institute of Computing Technology, Chinese Academy of Sciences. His research interests are in dynamic binary translation, multi-threaded program record-and-replay, operating systems, system virtualization, and memory corruption attacks and defenses.
\end{IEEEbiography}

\vfill

\end{document}